\documentclass[letterpaper,12pt]{article}
\usepackage{tabularx} 
\usepackage{amsmath}  
\usepackage{graphicx} 
\usepackage[margin=1in,letterpaper]{geometry} 
\usepackage{cite} 
\usepackage[final]{hyperref} 
\hypersetup{
	colorlinks=true,       
	linkcolor=blue,        
	citecolor=blue,        
	filecolor=magenta,     
	urlcolor=blue         
	}

\usepackage{float}

\newcounter{fig}\newcommand{\lbfig}[1]{\refstepcounter{fig}\label{#1}}

\allowdisplaybreaks 

\def\a{\alpha}

\def\r{\rho}

\def\m{\mu}
\def\n{\nu}

\def\s{\sigma}

\def\p{\pi}
\def\f{\phi}
\def\L{\Lambda}

\def\nn{\nonumber}
\def\2{\;\;}
\def\4{\;\;\;\;}

\def\beq{\begin{equation}}
\def\eeq{\end{equation}}
\def\bea{\begin{eqnarray}}
\def\eea{\end{eqnarray}}

\begin{document}

\begin{titlepage}

\vspace*{1cm}
\begin{center}
{\bf \Large On the Existence of Solutions with a Horizon\\[3mm]
in Pure Scalar-Gauss-Bonnet Theories }

\bigskip \bigskip \medskip

{\bf A. Bakopoulos}$^{\,(a)}$\,\footnote{Email: abakop@cc.uoi.gr},
{\bf P. Kanti}$^{\,(a)}$\,\footnote{Email: pkanti@uoi.gr} and
{\bf N. Pappas}$^{\,(b,c)}$\,\footnote{Email: npappas@uoi.gr}

\bigskip
$^{(a)}${\it Division of Theoretical Physics, Department of Physics,\\
University of Ioannina, Ioannina GR-45110, Greece}

\medskip
$^{(b)}${\it Nuclear and Particle Physics Section, Physics Department,\\ National and Kapodistrian
University of Athens, Athens GR-15771, Greece}

\medskip
$^{(c)}${\it Department of Physics, University of Thessaly, Lamia, GR-35100, Greece}

\bigskip \medskip
{\bf Abstract}
\end{center}
We consider the Einstein-scalar-Gauss-Bonnet theory and assume that,  at regimes of
large curvature, the Ricci scalar may be ignored compared to the quadratic Gauss-Bonnet
term. We then look for static, spherically-symmetric, regular black-hole solutions
with a non-trivial scalar field. Despite the use of a general form of the spacetime
line-element, no black-hole solutions are found. In contrast, solutions that resemble
irregular particle-like solutions or completely regular gravitational solutions with a finite
energy-momentum tensor do emerge. In addition, in the presence of a cosmological
constant, solutions with a horizon also emerge, however, the latter corresponds to a 
cosmological rather than to a black-hole horizon. It is found that, whereas the Ricci
term works towards the formation of the positively-curved topology of a black-hole
horizon, the Gauss-Bonnet term exerts a repulsive force that hinders the formation
of the black hole. Therefore, a pure scalar-Gauss-Bonnet theory cannot sustain any
black-hole solutions. However, it could give rise to interesting cosmological or
particle-like solutions where the Ricci scalar plays a less fundamental role.

\end{titlepage}

\setcounter{page}{1}

\section{Introduction}

After a century of searching and anticipation, at last, gravitational waves -- signals from 
processes taking place in strong-gravity regimes in our Universe -- have been successfully
detected \cite{LIGO, VIRGO}. This development has refueled the interest in the construction of
a more fundamental theory of Gravity. In most cases, such a theory includes extra fields or
higher-curvature terms in its action \cite{Stelle, General}, the presence of which modifies
the characteristics of the emergent gravitational solutions compared to the ones arising in the
context of the traditional General Relativity (GR). 

The quest for novel black-hole solutions in the context of a generalised gravitational theory
has been the most intense of all. The restrictive no-hair theorem \cite{NH-scalar} of GR, that
applied to gravitational theories including minimally-coupled scalar fields, was evaded when
novel black-hole solutions with Yang-Mills \cite{YM}, Skyrme fields \cite{Skyrmions} or
fields with a conformal coupling to gravity \cite{Conformal} appeared in the literature. A
novel formulation of the no-hair theorem \cite{Bekenstein} was also evaded in the context
of a gravitational theory with a scalar field coupled to the Gauss-Bonnet term, a quadratic
curvature term \cite{Zwiebach, Gross, Metsaev}: the discovery of the dilatonic black holes
\cite{DBH} (see also \cite{Gibbons, Callan, Campbell, Mignemi, Kanti1995} for some earlier
studies) was soon followed by the one of the coloured black holes  \cite{Torii, KT}, in the
presence of a Yang-Mills field, and then of higher-dimensional \cite{Guo} or rotating
versions \cite{Kleihaus, Pani, Herdeiro, Ayzenberg} (see \cite{Win-review, Charmousis-rev,
Herdeiro-review, Blazquez} for a number of reviews).

After a dormant period, the revival of the Horndeski \cite{Horndeski} and Galileon \cite{Galileon}
theories gave a significant boost to the concept of generalised gravitational theories, that
contain a single scalar field and higher-derivative curvature terms. Even the no-hair theorems
were re-formulated \cite{SF, HN} but to no avail: novel black-hole solutions were again
constructed \cite{SZ, Babichev, Benkel, Yunes2011}. These solutions, as well as the earlier ones
mentioned above, have the characteristic feature of the scalar hair: a regular, non-trivial
scalar field that is associated with the black hole, a feature forbidden by GR. In a recent
work \cite{ABK1}, it was demonstrated that a general class of theories with a scalar field
having an arbitrary coupling function to the quadratic Gauss-Bonnet term, the
Einstein-scalar-Gauss-Bonnet (EsGB) class of theories, always evades
the no-hair theorem \cite{Bekenstein} and leads to regular, black-hole solutions with
scalar hair. The early dilatonic \cite{DBH} and shift-symmetric Galileon \cite{Benkel} solutions
are particular examples of this general statement -- as are also the additional solutions
\cite{Doneva, Silva} that appeared almost simultaneously with \cite{ABK1}. 
In addition, a large number of works has appeared that studied novel black holes or
compact objects in these, or similar, types of theories as well as their properties
\cite{Bardoux}-\cite{Barrientos}.
The asymptotically-flat black-hole solutions were also supplemented by solutions with
an asymptotic (Anti)-de Sitter behaviour, a topic that has also attracted a lot of interest
in the literature \cite{Martinez-deSitter}-\cite{Radu-scal-dS}.

The Einstein-scalar-Gauss-Bonnet theory has in fact proven to be an extremely rich generalised
theory of gravity. Apart from novel black-hole solutions as described above, it has been shown
to lead to families of wormholes that require no exotic matter \cite{KKK1, ABKKK} and 
particle-like solutions with regular spacetimes \cite{Hartmann, KKK2}, all with non-trivial
scalar hair (see also \cite{Herdeiro-Oliveira, Afonso, Radu-part, Canate1, Canate2}). The presence
of the quadratic Gauss-Bonnet term seems to be of paramount importance for the emergence
of all of these compact solutions. It creates an effective energy-momentum tensor that may
locally violate the energy conditions while the actual matter fields of the theory continue to
respect them. This leads to the evasion of non-existence arguments of GR and the emergence
of novel solutions, from black holes to wormholes, with a scalar hair. 

The Einstein-scalar-Gauss-Bonnet theory has also interesting cosmological implications,
and these were in fact the first ones to be studied in the literature. In the context of the
effective heterotic superstring theory, where the scalar field was identified with one of the
moduli fields of the theory, it was shown that this theory leads to singularity-free
cosmological solutions \cite{ART}. Later, it was demonstrated that similar type of solutions
emerge for a variety of coupling functions, with a number of common features, between
the scalar field and the Gauss-Bonnet term \cite{RT, KRT}. More recently, the same theory
was studied from a novel perspective \cite{KGD}: as we go backwards in time, the curvature of
spacetime considerably increases and the quadratic Gauss-Bonnet term becomes eventually as
important as, or even larger than, the linear Ricci term. Assuming that such a time period
exists, the Ricci term was altogether ignored from the theory and the coupled system
of the scalar field and the Gauss-Bonnet was studied on its own. It was found \cite{KGD}
that this simplified theory supported singularity-free solutions with the same characteristics
as the ones emerging in the context of the complete theory. In addition, a family of
attractive inflationary solutions with a natural exit mechanism was found, all in an
analytical way due to the simplification of the set of field equations.

In the context of the present work, we will keep the same perspective but focus on the
emergence of solutions with a horizon. Our main priority will be to investigate whether
the simplified, pure scalar-Gauss-Bonnet theory leads to regular black-hole solutions
with a non-trivial scalar field. We will therefore assume that there exists a part of
spacetime where the Gauss-Bonnet term may dominate over the linear Ricci term. We will
seek for static, spherically-symmetric, regular black holes, and attempt to solve the
simplified set of field equations by using both analytical and numerical methods. We will
also use alternative forms of line-elements in an effort to increase the flexibility of our
ansatz. The presence of a cosmological constant will also be employed, and its role
to the formation of a horizon will be investigated. As we will demonstrate, the pure
scalar-Gauss-Bonnet theory cannot support by itself black-hole spacetimes -- this
will be due to the conflicting roles of the Gauss-Bonnet and Ricci terms in the formation
of a regular black-hole horizon. Nevertheless, our quest for black-hole solutions
will lead us instead to a number of alternative solutions -- all parts of the phase-space
of solutions of the pure scalar-Gauss-Bonnet theory -- with a number of interesting
characteristics. 

The outline of the present work is as follows: in Section 2, we present our theoretical framework
and field equations. In Section 3, we look, in an analytical way, for solutions with a horizon 
in the case where the cosmological constant vanishes -- we also perform an exact
numerical study of a family of regular solutions we derive. In Section 4, we employ a
generalised form of the spacetime line-element, and repeat our previous analysis. We
re-instate the cosmological constant in Section 5, and interpret the solutions we obtain.
In Section 6, we use exact numerical results for black-hole solutions found in the
context of the complete Einstein-scalar-Gauss-Bonnet theory, and investigate the
role of the Gauss-Bonnet and Ricci terms in the formation of a black-hole
horizon. We finish with our conclusions in Section 7.

%

\section{The Theoretical Framework}

The starting point of our analysis will be the following action, describing a generalised theory of gravity
\begin{equation}\label{action}
S=\frac{1}{16\pi}\int{d^4x \sqrt{-g}\left[R-\frac{1}{2}\,\partial_{\mu}\phi\,\partial^{\mu}\phi+
f(\phi)R^2_{GB} - 2\L \right]}.
\end{equation}
The theory contains the Ricci scalar curvature $R$, a scalar field $\phi$ and the higher-curvature,
quadratic Gauss-Bonnet (GB) term defined as
\begin{equation}\label{GB}
R^2_{GB}=R_{\mu\nu\rho\sigma}R^{\mu\nu\rho\sigma}-4R_{\mu\nu}R^{\mu\nu}+R^2,
\end{equation}
in terms of the Riemann tensor $R_{\mu\nu\rho\sigma}$, Ricci tensor $R_{\mu\nu}$ and Ricci
scalar $R$. The GB term, being a topological invariant in four dimensions, must be coupled
to the scalar field $\phi$. This is realised via the arbitrary coupling function $f(\phi)$; choosing 
different forms for the coupling function, one may study the emergence of solutions within
a whole {\it class} of theories. The theory includes also a cosmological constant $\Lambda$.
Throughout our work, we will use units in which $G=c=1$.

Taking the variation of the action with respect to the metric $g_{\mu\nu}$ and the scalar field
$\phi$, we end up with the gravitational field equations and the equation of motion for the scalar
field. These have the following forms
\begin{equation}
G_{\mu\nu}=T_{\mu\nu}\,, \label{field-eqs}
\end{equation}
\begin{equation}
\nabla^2 \phi+\dot{f}(\phi)R^2_{GB}=0\,, \label{phi-eq_0}
\end{equation}
respectively. In the above, $G_{\mu\nu}$ is the Einstein tensor and $T_{\mu\nu}$ is the
total energy-momentum tensor of the theory
\begin{equation}\label{Tmn}
T_{\mu\nu}=-\frac{1}{4}\,g_{\mu\nu}\,\partial_{\rho}\phi\,\partial^{\rho}\phi+
\frac{1}{2}\,\partial_{\mu}\phi\,\partial_{\nu}\phi-\frac{1}{2}\left(g_{\rho\mu}g_{\lambda\nu}+g_{\lambda\mu}g_{\rho\nu}\right)
\eta^{\kappa\lambda\alpha\beta}\tilde{R}^{\rho\gamma}_{\quad\alpha\beta}
\nabla_{\gamma}\partial_{\kappa}f(\phi)- \L\,g_{\m\n}\,,
\end{equation}
that receives contributions from the kinetic term of the scalar field, the GB term and
the cosmological constant. The dot over the coupling function denotes its derivative with
respect to the scalar field (i.e. $\dot f =df/d\phi$). We have also used the definition
\begin{equation}\label{tildeR}
\tilde{R}^{\rho\gamma}_{\quad\alpha\beta}\equiv\eta^{\rho\gamma\sigma\tau}
R_{\sigma\tau\alpha\beta}\equiv\frac{\epsilon^{\rho\gamma\sigma\tau}}{\sqrt{-g}}\,
R_{\sigma\tau\alpha\beta}\,.
\end{equation}

The emergence of regular black-hole solutions with a non-trivial scalar field, in the
context of the theory (\ref{action}) and for a variety of coupling functions $f(\phi)$,
was demonstrated in \cite{ABK1, BAK} with either Minkowski or Anti-de Sitter asymptotic
behaviour.
Here, we will investigate whether regular, black-hole solutions with scalar hair emerge
in the context of the pure scalar-GB theory, i.e. in the absence of the Ricci scalar from
the theory. In that case, the derived solutions would rely solely on the synergy between
the scalar field and the GB term. From the field equations (\ref{field-eqs})-(\ref{phi-eq_0}),
we may see that such a synergy is in principle possible: a non-trivial scalar field ensures
the presence of the GB term in the theory whereas the GB term provides in its turn a
non-trivial potential for the scalar field. It is this same synergy that leads to
singularity-free or inflationary cosmological solutions, even in the absence of the
Ricci scalar from the theory, as was analytically demonstrated in \cite{KGD}.  

The assumption that the linear gravitational Ricci term may be ignored from the theory
when compared to the quadratic GB term may be justified only in regimes of spacetime
where the curvature is particularly large. This may be realised only near the black-hole
horizon whereas in the asymptotic regime the Ricci scalar must be necessarily re-instated.
Therefore, the question we would like to pose, and investigate in what follows, is the
following: does the curvature of spacetime ever become so strong that a black-hole horizon
may be formed only due to the effect of the GB term (supplemented by that of the scalar field)?
  
To this end, we will assume a static, spherically-symmetric ansatz for the spacetime
line-element of the form:
\begin{equation}\label{metric}
{ds}^2=-e^{A(r)}{dt}^2+e^{B(r)}{dr}^2+r^2({d\theta}^2+\sin^2\theta\,d\varphi^2).
\end{equation}
In accordance to the above discussion, we will focus on the ``near'' regime of spacetime,
i.e. on the small-$r$ regime, and assume that, there, all terms associated with the Ricci 
term may be ignored from the field equations. That amounts to ignoring altogether the
components of the Einstein tensor $G_{\mu\nu}$ from the gravitational field equations
(\ref{field-eqs}). Then, employing the ansatz (\ref{metric}), the explicit form of the
components of Einstein's equations becomes
\begin{align}
T^t_{\;\,t}=&-\frac{e^{-2B}}{4r^2}\left[\phi'^2\left(r^2e^B+16\ddot{f}(e^B-1)\right)-8\dot{f}\left(B'\phi'(e^B-3)-2\phi''(e^B-1)\right)\right]-\L=0, \label{Ttt}\\[2mm]
T^r_{\;\,r}=&\frac{e^{-B}\phi'}{4}\left[\phi'-\frac{8e^{-B}\left(e^B-3\right)\dot{f}A'}{r^2}\right] -\L=0, \label{Trr}\\[0mm]
T^{\theta}_{\;\,\theta}=&T^{\varphi}_{\;\,\varphi}=-\frac{e^{-2B}}{4 r}\left[\phi'^2\left(re^B-8\ddot{f}A'\right)-4\dot{f}\left(A'^2\phi'+2\phi'A''+A'(2\phi''-3B'\phi')\right)\right]
-\L=0. \label{Tthth}
\end{align}
We observe that, upon ignoring the components of the Einstein tensor from the field equations,
the total {\it effective} energy-momentum tensor vanishes. However, this is due not to the
triviality of the matter distribution in our theory but to the cancellation of the positive
contribution of the kinetic term of the scalar field and the negative contribution of the
GB term to the {\it effective} energy-density and pressure components of the system. 
We therefore look for non-trivial configurations of the scalar field and metric functions
that satisfy the above equations and thus correspond to {\it locally} zero-energy and
zero-pressure solutions. 

The scalar field equation (\ref{phi-eq_0}) on the other hand remains unaltered by the
elimination of the Ricci scalar in the small-$r$ regime, and assumes the following explicit
form:
\begin{equation}
2r\phi''+(4+rA'-rB')\,\phi'+\frac{4\dot{f}e^{-B}}{r}\left[(e^B-3)A'B'-(e^B-1)(2A''+A'^2)\right]=0\,, \label{phi-eq}
\end{equation}
where we have assumed that the scalar field shares the symmetry of spacetime and thus is
also static and spherically-symmetric, $\phi=\phi(r)$. In all the above equations, the prime
denotes differentiation with respect to the radial coordinate $r$.

%
%

\section{\texorpdfstring{Solutions in the ``near'' regime for $\Lambda=0$}{Solutions in the ``near'' regime for L=0}}
The issue of the emergence of regular, black-hole solutions from the system of
Eqs. (\ref{Ttt})-(\ref{phi-eq}), with $\Lambda=0$, was briefly discussed in \cite{ABK1}.
Here, we will first review and expand on the mathematical arguments involved
in that analysis and, second, study and characterise the families of solutions we obtain. 
We will again assume that the cosmological constant is zero and postpone the study
of its role for a later section.

As our priority is to find solutions with a horizon, we will demand that, for some value
of the radial coordinate $r=r_h$, the following conditions should hold
\beq
g_{tt}\,|_{r=r_h^+}\rightarrow 0\,, \qquad \quad 
g_{rr}\,|_{r=r_h^+}\rightarrow \infty\,. \label{metric-cond}
\eeq
The above conditions amount to assuming that $A'\rightarrow \infty$ and
$B'\rightarrow \infty$. In fact, one may consider any of the two conditions as
the starting point of the analysis -- in the case of the emergence of a 
spherically-symmetric black hole, these two conditions are equivalent; however,
in the absence of such a solution, each condition allows us to explore different
parts of the phase space of the solutions of the theory. We investigate these two
different lines of thinking in the following two subsections.

 
\subsection{\texorpdfstring{Expanding Around $A'\rightarrow \infty$ }{Expanding Around  A'}}

Before applying any limit, we may observe that Eq. (\ref{Trr}) can be solved to yield an
expression for the metric component $e^B$, that is
\begin{equation}\label{B-fun}
e^{B}=\frac{24 A' \dot{f}}{8 A' \dot{f}-r^2 \phi '}\,,
\end{equation}
from which we may easily deduce an expression for the first derivative of $B$, namely
\begin{equation}
B'=\frac{r \left[A' \left(\dot{f} \left(r \phi ''+2\phi '\right)-r \phi '^2 \ddot{f}\right)-r
   A'' \phi ' \dot{f}\right]}{A' \dot{f}\left(8 A' \dot{f}-r^2 \phi '\right)}\,.\label{B-prime}
\end{equation}
Employing Eqs. (\ref{B-fun}) and (\ref{B-prime}), we may eliminate the metric function
$B(r)$ and its derivative from the remaining field equations (\ref{Ttt}), (\ref{Tthth}) and
(\ref{phi-eq}). The latter reduce to a set of two independent second-order, coupled,
ordinary differential equations that may be brought to the following form
\beq
A''= \frac{{P}}{{S}}\,, \qquad \quad 
\phi''=\frac{{Q}}{{S}}\,, \label{Aphi-sys}
\eeq
where:
\beq
{S}=16 e^{B} \dot{f} \left(16 A' \dot{f}-r^2 \phi'\right)+16 \dot{f} \left(5 r^2 \phi '
    -16 A'\dot{f}\right),\label{S-def}
\eeq
\begin{align}
{P}=&e^{2 B} \left(32 r A' \phi ' \dot{f}-6 r^3 \phi'^2\right)+e^{B} \left(-64 r^2 A' \phi '^2
   \ddot{f}+40 r^2 A'^2 \phi ' \dot{f}+32 r A' \phi ' \dot{f}\right.\nonumber\\
   &\left.-128 A'^3 \dot{f}^2-5 r^4 A' \phi '^2+10 r^3 \phi '^2\right)-40 r^2
   A'^2 \phi ' \dot{f}+128 A'^3 \dot{f}^2,\\[2mm]
{Q}=&-\frac{1}{A'}\left[e^{B} \phi '^2 \left(16 r^2 A' \phi' \ddot{f}+8 r^2 A'^2 \dot{f}-
32 r A' \dot{f}+256 A'^2 \dot{f} \ddot{f}+r^4 A' \phi '-6r^3 \phi '\right)\right.\nonumber\\
   &\left.+\phi '^2 \left(-16 r^2 A' \phi' \ddot{f}+24 r^2 A'^2 \dot{f}+96 r A'
  \dot{f}-256 A'^2 \dot{f} \ddot{f}\right)+2 r^3 e^{2 B} \phi '^3\right].\label{Q-def}  
\end{align}
In the above expressions, we have eliminated $B'$ but, for simplicity of notation, we kept
$e^B$ -- the latter should be considered as a dependent function of the remaining
independent variables according to Eq. (\ref{B-fun}).  

We will now assume that, as $r \rightarrow r_h$, $A'$ diverges. The regularity of the
horizon, if existent, demands that the scalar field $\phi$ as well as its first and second
derivative remain there finite. Under these assumptions, Eq. (\ref{B-fun}) readily gives
that
\begin{equation}\label{94}
e^B=3+\mathcal{O}\left(\frac{1}{A'}\right)\,.
\end{equation}
By replacing the above in Eqs. (\ref{Aphi-sys})-(\ref{Q-def}), we find that, near the point of
interest, the following relations hold
\beq
A''=-\frac{1}{2}A'^2+\mathcal{O}\left(A'\right)\,, \qquad 
\phi''=\;\mathcal{O}(1)\,.\label{96}
\eeq
The above equations may be easily integrated with respect to the radial coordinate to give:
\begin{equation}
A'=\frac{2}{r-r_h}+\mathcal{O}(1)\,, \qquad \phi'=\phi_1 (r-r_h) + \mathcal{O}(1)\,.
\end{equation}
The first of the above equations consistently gives that, near $r_h$, $A'\rightarrow \infty$ as
assumed above. Integrating once more, we find the complete form of the asymptotic solution
for the metric functions $A$, $B$  and the scalar field $\phi$ near $r_h$:
\begin{align}
e^A&=a_2\,(r-r_h)^2+\mathcal{O}\left((r-r_h)^3\right),\\[2mm]
e^B&=3+b_1(r-r_h)+\mathcal{O}\left((r-r_h)^2\right),\label{exppbb}\\[2mm]
\phi&=\phi_0+\phi_1(r-r_h)+\phi_2(r-r_h)^2+\mathcal{O}\left((r-r_h)^3\right).
\label{exp-Aphi}
\end{align}
In the above, $b_1$, $a_2$ and $\phi_i$ are arbitrary integration constants. As demanded, the scalar field
and its derivatives are regular at $r_h$ while the metric function $e^A$ vanishes thus exhibiting
the expected behaviour near a black-hole horizon. However, the behaviour of the second metric
function $e^B$, as given by Eq. (\ref{exppbb}), does not describe a black hole as it remains regular
near $r_h$. In order to gain more information about the form of spacetime around $r_h$, we
calculate the scalar gravitational quantities, the exact expressions of which may be found in the
Appendix \ref{scalar}. We then find the following results:
\begin{align}
R&= -\frac{10b_1}{3(r-r_h)} +\mathcal{O}(1),   \\[1mm] 
R_{\m\n}R^{\m\n}&=  \frac{50b_1}{9(r-r_h)^2} +\mathcal{O}\left( \frac{1}{r-r_h}\right) ,    \\[1mm] 
R_{\m\n\r\s}R^{\m\n\r\s}&= \frac{100b_1}{9(r-r_h)^2} +\mathcal{O}\left( \frac{1}{r-r_h}\right),\\[1mm]
R^2_{GB}&= -\frac{40b_1}{3r_h^2(r-r_h)}+ \mathcal{O}\left( 1\right)\,.
\end{align}
According to the above, all curvature invariant quantities  $R$, $R_{\m\n}R^{\m\n}$ and $R_{\m\n\r\s}R^{\m\n\r\s}$ are diverging near $r_h$ while the Gauss-Bonnet combination
exhibits a softer divergence than expected as the dominant terms of order $(r-r_h)^{-2}$
exactly cancel. 

Under the change of coordinate $l=r-r_h$, the expansions (\ref{exp-Aphi}) resemble the
asymptotic form of a particle-like solution near the origin. As discussed above, our solution
is characterised
by a spacetime singularity that, at first, may be considered as unphysical. However, particle-like
solutions plagued by singularities, either in spacetime or in the profile of the scalar field,
are quite common in the literature in the context of scalar-tensor theories of gravity
(see, for instance \cite{KKK2, Afonso}). These solutions are physical whenever they are
characterised by a finite total energy-momentum tensor as is the case also for the
solutions derived here.


\subsection{\texorpdfstring{Expanding Around $B'\rightarrow \infty$ }{Expanding Around  B'}}\label{subse1}

Alternatively, we may employ the fact that near a black-hole horizon it holds that
$e^B \rightarrow \infty$. As we will shortly confirm, this amounts to assuming that
$B'\rightarrow \infty$. The metric function $A$ will now be considered as a dependent
variable, and Eq. (\ref{Trr}) may be readily solved to give:
\begin{equation}\label{99}
A'=\frac{r^2 e^{B} \phi '}{8 \left(e^{B}-3\right) \dot{f}}\,.
\end{equation}
Computing also the second derivative of $A$ from the above expression, we may
eliminate the metric function $A$ and its derivatives from the remaining field equations. Then,
we form a system of two coupled, ordinary differential equations, one first-order and one
second-order, for the metric function $B$ and the scalar field $\phi$, respectively. 
These have the form:
\beq
B'=\frac{\mathcal{Y}}{\mathcal{W}}\,, \qquad \quad 
\phi''=\frac{\mathcal{X}}{\mathcal{W}}\,,\label{Bphi-sys}
\eeq
where
\begin{align}
\mathcal{W}=&32 A' \dot{f}\,,\label{102}\\\nonumber\\
\mathcal{Y}=&2 e^{-B} \left[-2 \left(e^{B}-1\right) \left(r e^{B} \phi '-8 A''
   \dot{f}\right)+8 \left(e^{B}-1\right) A'^2 \dot{f}+r^2
   \left(-e^{B}\right) A' \phi '\right],\\\nonumber\\
\mathcal{X}=&2e^{-B}\phi' \left(e^{B}-3\right) \left(r e^{B} \phi '-8 A'' \dot{f}\right)+e^{B} A' \phi ' \left(32 \ddot{f}+3 r^2\right)-8
   \left(e^{B}-3\right) A'^2 \dot{f}\,.\label{104}
\end{align}
Again, for notational simplicity we have kept $A'$ and $A''$ in the expressions above, however,
these quantities are now dependent functions of the independent variables $B$ and $\phi$. 

Let us now focus on the regime near $r_h$ where $e^B$ and $B'$ are assumed to diverge.
First, we expand there Eq. (\ref{99}) to obtain:
\begin{equation}
A'=\frac{r^2\phi'}{8\dot{f}}+\mathcal{O}\left(e^{-B}\right). \label{A'-appr}
\end{equation}
Substituting the above result into Eqs. (\ref{Bphi-sys})-(\ref{104}), we find that near $r_h$
the following relations hold:
\begin{align}
B'=&-\frac{2}{r}\,e^B+\mathcal{O}\left(e^{-B}\right),\label{106}\\
\phi''=&-\frac{e^B}{r}\,\phi'+\mathcal{O}\left(e^{-B}\right).\label{107}
\end{align}
According to Eq. (\ref{106}), near $r_h$, $B'$ indeed diverges as assumed. If we integrate
this equation with respect to the radial coordinate,  we find that:
\begin{equation}
e^{-B}=2\ln\left(\frac{r}{r_h}\right). \label{B-sol}
\end{equation}
Indeed, as demanded, the metric function $e^B$ exhibits the desired behaviour near a black-hole
horizon. In order for this horizon to be also regular, the scalar field and its derivatives should
be finite. Then, Eq. (\ref{107}) dictates that we must necessarily have $\phi'(r_h)=0$. 
We may in fact solve analytically Eqs. (\ref{A'-appr}) and (\ref{107}) to obtain the solutions
\beq
\phi=\phi_0+\phi_1 \left(-\frac{1}{2}\sqrt{\p}r_h\, 
\text{Erfi}\left(\xi\right)+r\, \xi \right),\qquad 
\phi'=\phi_1\, \xi,\label{aa4}
\eeq
and
\beq
A= a_0-\frac{r^3\phi_1\left( \sqrt{3}\, F_D\left( \sqrt{3}\,\xi \right)   -3\,\xi \right)}{72 \dot f(\phi_0)},
\eeq
where $a_0$ and $\phi_i$ are again integration constants, and where we have defined the variable
\beq
\xi=\sqrt{\ln\left(\frac{r}{r_h}\right)}. \label{xi}
\eeq
Also, $F_D(x)$ and $\text{Erfi}(x)$ are the Dawson and error function, respectively, defined as:
\beq
F_D(x)=e^{-x^2}\int_0^xe^{t^2}dt =\frac{\sqrt{\p}}{2}e^{-x^2}\text{Erfi}(x).
\eeq

Although the solution for the metric function $e^B$ (\ref{B-sol}) hints to the presence
of a horizon, on which the scalar field remains regular according to Eqs. (\ref{aa4}),
the behaviour of the metric function $e^A$ reveals that this solution is not a black hole:
in the limit $r \rightarrow r_h$, or $\xi \rightarrow 0$, $A'$ remains finite and $A$
adopts an arbitrary constant value. Using the above asymptotic solutions, we may
calculate once again the scalar curvature quantities, the expressions of which are
listed below:
\begin{align}
R&= -\frac{2}{r_h^2} +\mathcal{O}(\sqrt{r-r_h}),   \\[2mm] 
R_{\m\n}R^{\m\n}&=  \frac{4}{r_h^4} +\mathcal{O}(\sqrt{r-r_h}),    \\[2mm] 
R_{\m\n\r\s}R^{\m\n\r\s}&= \frac{12}{r_h^4} +\mathcal{O}(\sqrt{r-r_h}), \\[2mm]
R^2_{GB}&=-\frac{\phi_1\sqrt{r-r_h}}{r_h^{3/2}\dot f(\f_0)}+\mathcal{O}\left( (r-r_h)^{3/2} \right).
\end{align}
The above asymptotic values reveal that, near $r_h$, the spacetime remains regular
and no singularities emerge. All curvature invariants assume constant values apart
from the GB combination that is vanishing at exactly $r=r_h$. 

\begin{figure}[t]
\lbfig{metric} 
\begin{center}
\hspace{0.0cm} \hspace{-0.6cm}
\includegraphics[height=.23\textheight, angle =0]{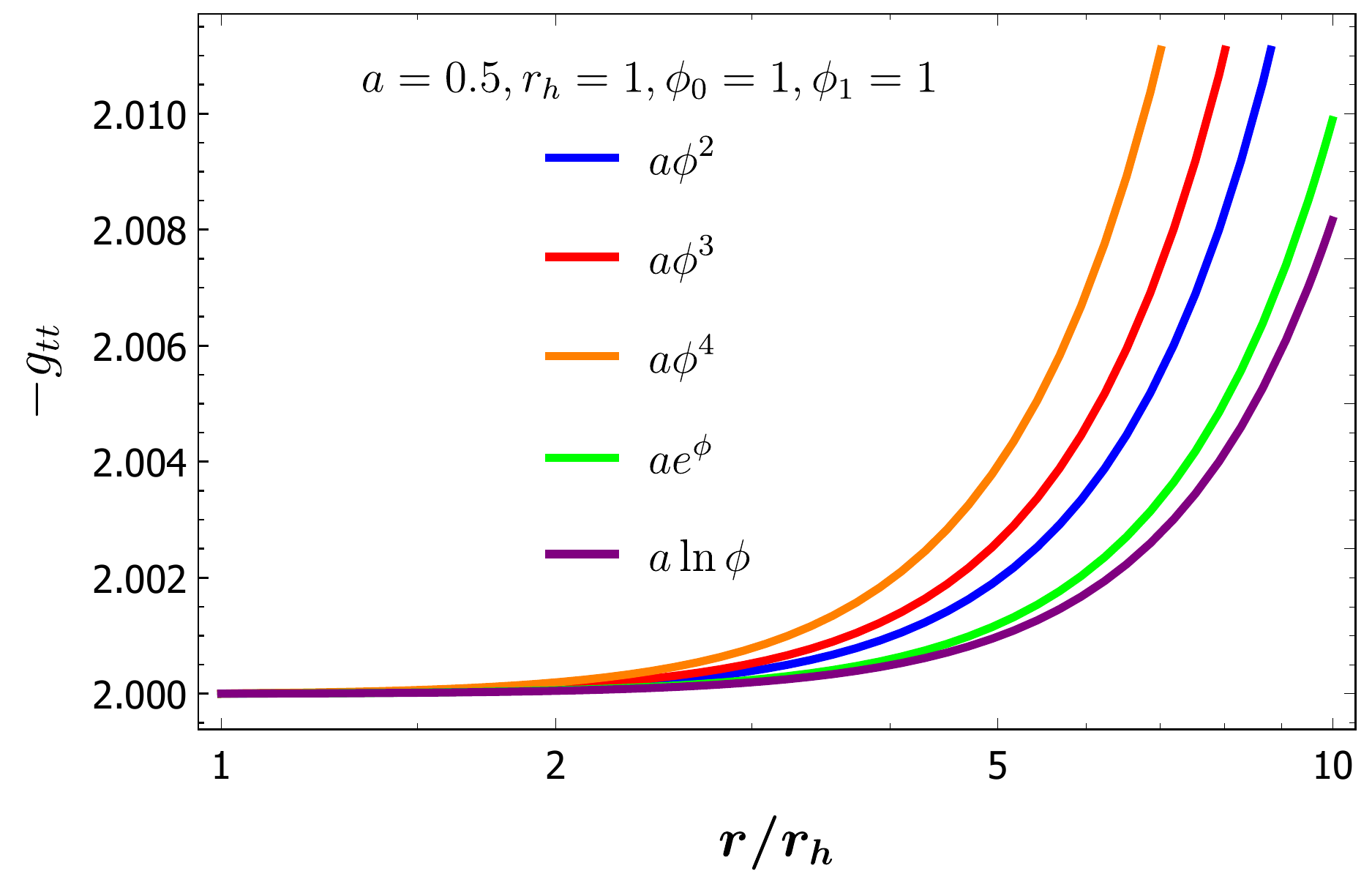}
\hspace{0.5cm} \hspace{-0.6cm}
\includegraphics[height=.23\textheight, angle =0]{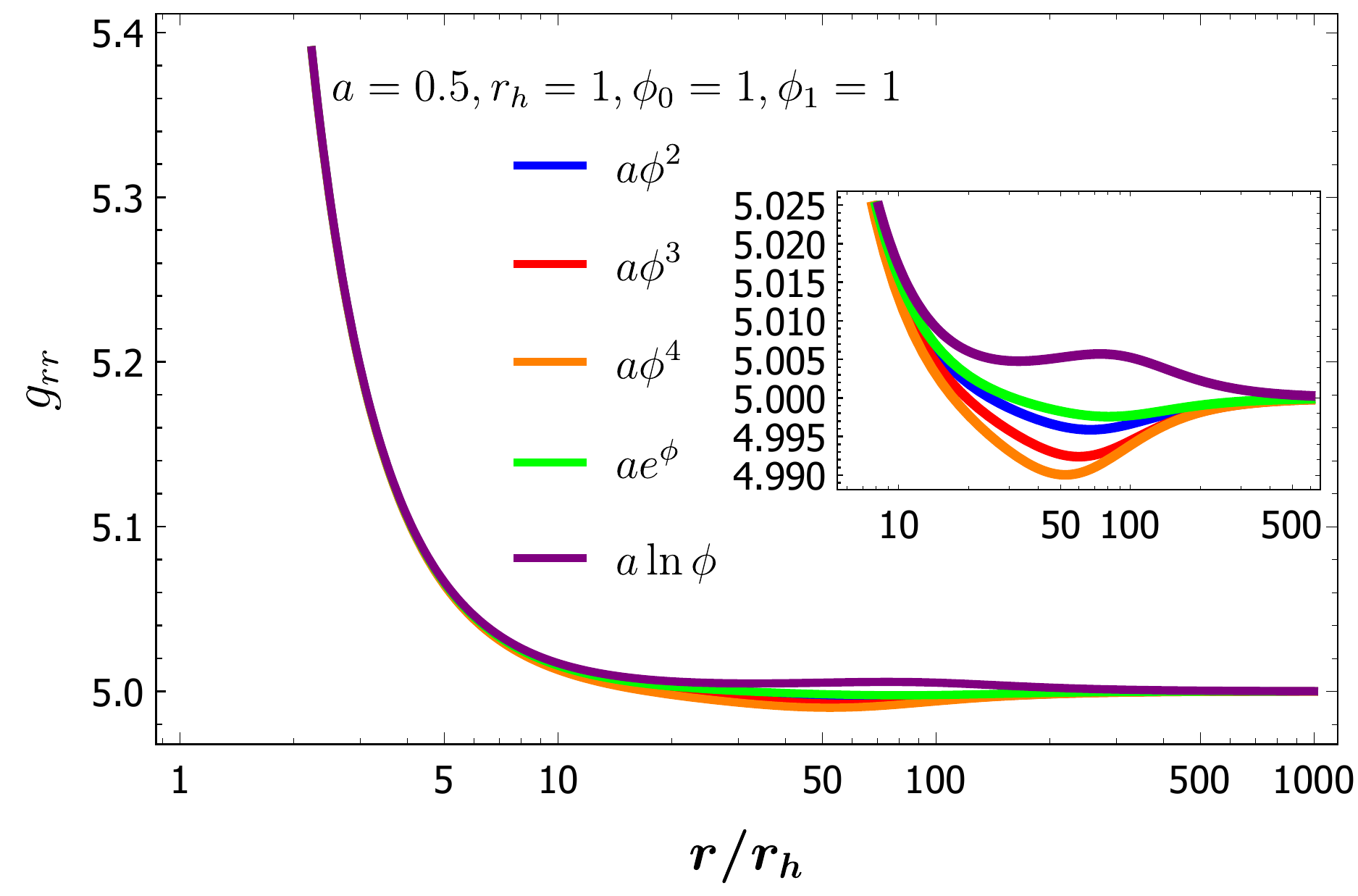}
\\
\hspace*{0.7cm} {(a)} \hspace*{7.5cm} {(b)}  \vspace*{-0.5cm}
\end{center}
\caption{
(a) The $-g_{tt}$ component, and (b) the $g_{rr}$ component of the metric tensor
for a pure scalar-Gauss-Bonnet solution and for a variety of forms of the coupling
function $f(\phi)$.}
\end{figure}

\begin{figure}[b!]
\lbfig{GB} 
\begin{center}
\includegraphics[height=.22\textheight, angle =0]{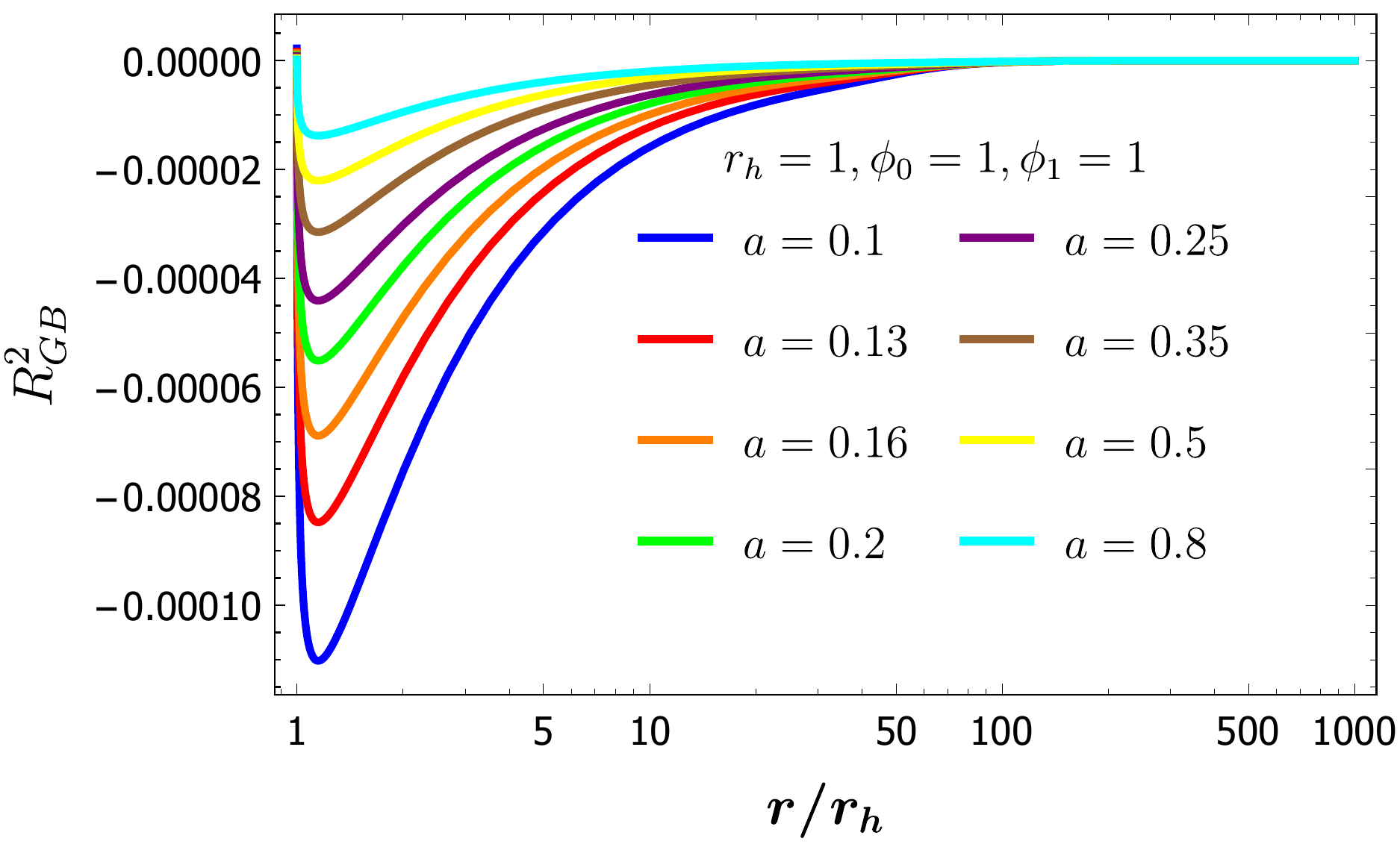}
\end{center}
\vspace*{-0.5cm}
\caption{
The Gauss-Bonnet term $R^2_{GB}$ for a pure scalar-Gauss-Bonnet
solution and a linear coupling function $f(\phi)=a\phi$, for
various values of $a$.}
\end{figure}

We have performed a numerical integration of the system (\ref{Bphi-sys}) to determine
the solutions for the metric function $B$ and scalar field $\phi$ in the whole radial regime. 
To this end, we have used Eqs. (\ref{B-sol}) and (\ref{aa4}) as boundary conditions,
and finally integrated Eq. (\ref{99}) with a randomly chosen boundary condition $e^{A(r_h)}=2$.
In Fig. \ref{metric}(a,b), we depict the profiles of the two metric components $-g_{tt}$
and $g_{rr}$, respectively, for a variety of forms of the coupling function $f(\phi)$.
We observe that the qualitative behaviour of the two metric components remains
largely unaffected by the exact form of $f(\phi)$, especially at the small-$r$ regime.
As our analytic calculations revealed, the $g_{rr}$ component diverges as
$r \rightarrow r_h$ but the $g_{tt}$ remains finite thus failing to adopt a black-hole
profile. The regularity of the spacetime is reflected in the finiteness of the GB term
depicted in Fig. \ref{GB}: this scalar gravitational quantity vanishes very close to and
far away from $r_h$ while having non-trivial values at finite, intermediate distance.

\begin{figure}[t!]
\lbfig{phi} 
\begin{center}
\hspace{0.0cm} \hspace{-0.6cm}
\includegraphics[height=.22\textheight, angle =0]{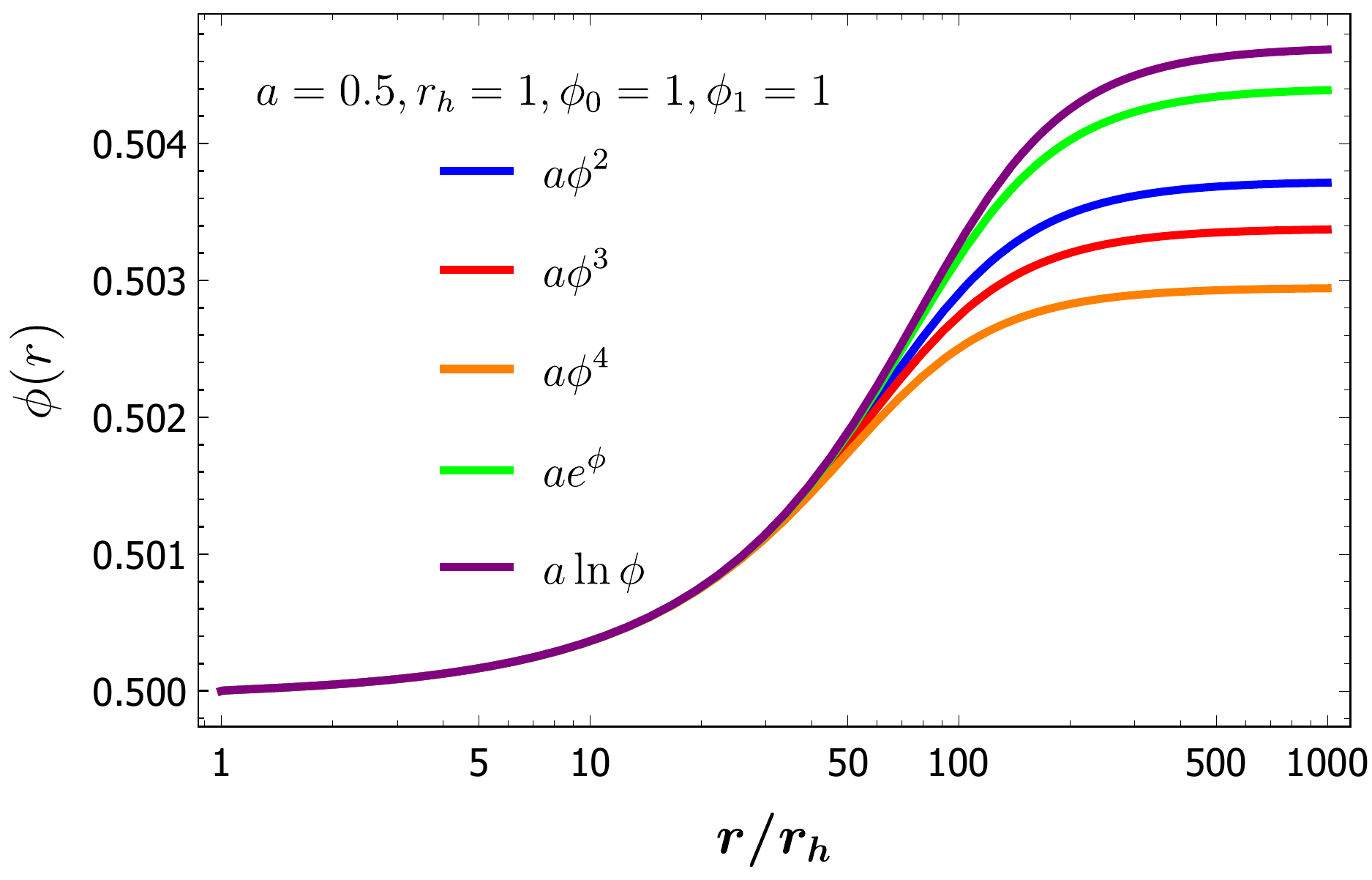}
\hspace{0.5cm} \hspace{-0.6cm}
\includegraphics[height=.22\textheight, angle =0]{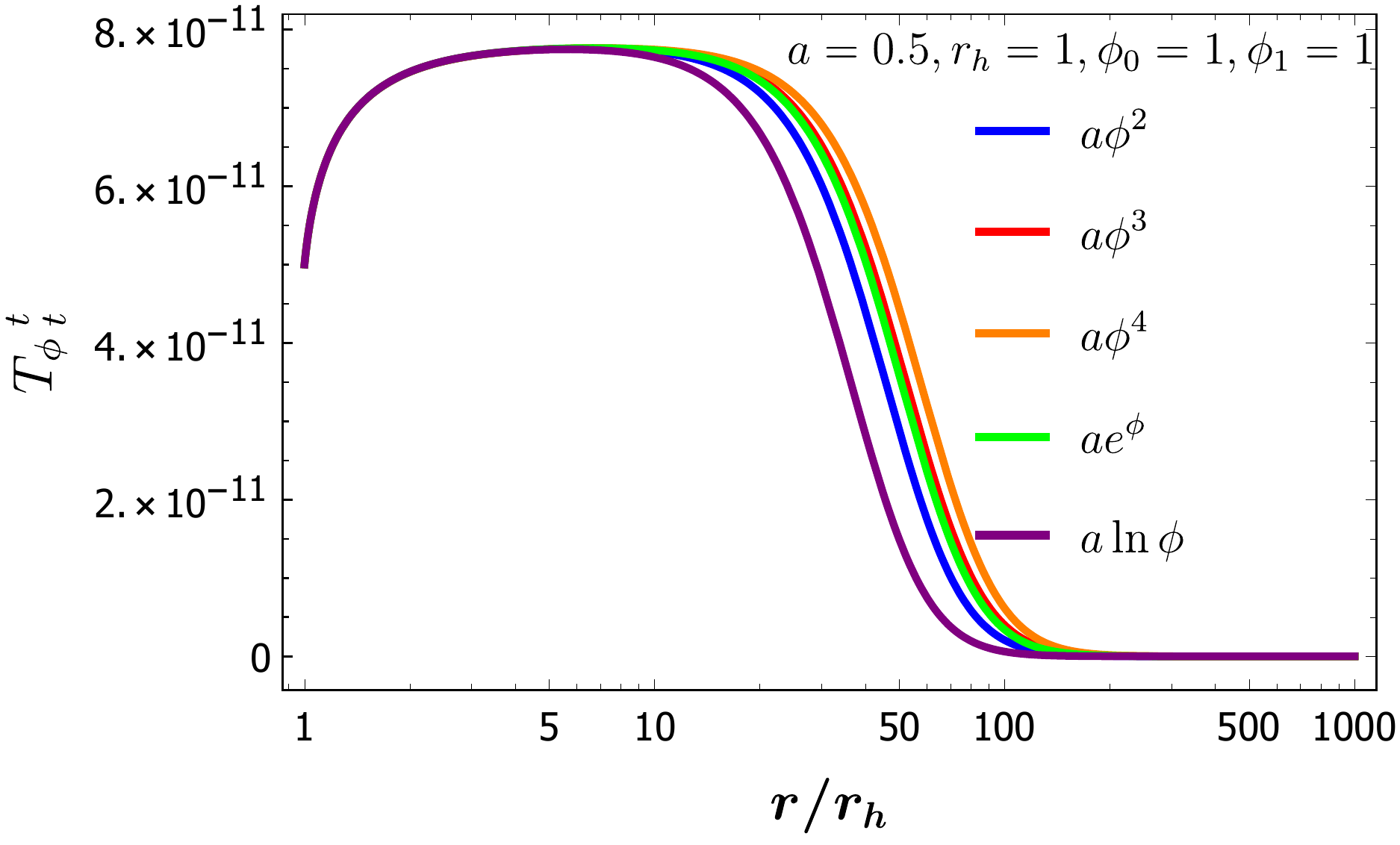}
\\
\hspace*{0.7cm} {(a)} \hspace*{7.5cm} {(b)}  \vspace*{-0.5cm}
\end{center}
\caption{
(a) The scalar field $\phi$ and (b) the $(tt)$-component of the scalar kinetic-term contribution
to the effective energy-momentum tensor, for various forms of the coupling function.}
\end{figure}

The solution for the scalar field $\phi$ is presented in Fig. \ref{phi}(a). As clearly shown,
it adopts finite constant values near $r_h$ and at asymptotic infinity which leads to its first
derivative having a vanishing value at both these regimes. This profile results into the form
of the contribution of the scalar kinetic-term to the effective energy-momentum tensor
that is shown in Fig. \ref{phi}(b): as expected, this contribution has a non-zero value
at intermediate values of the radial coordinate where the scalar field has a non-trivial
profile. The same is true for the contribution of the GB term to the total energy-momentum
tensor (\ref{Tmn}): this is again non-trivial at intermediate distances, in accordance to
Fig. \ref{GB}, and has exactly the same form as the $T_{\phi\,\,t}^{\ \,t}$ contribution,
depicted in Fig. \ref{phi}(b), but with the exact opposite sign: it is this behaviour that guarantees
the vanishing total energy-momentum tensor as our analysis demanded.
Let us stress that both families of solutions derived in this section correspond to non-trivial
gravitational solutions with finite, zero total energy-momentum tensor whose physical
significance will be studied elsewhere.

%
%

\section{Alternative Spacetime Line-elements}

The main objective of this work is to find an asymptotic solution that describes a regular
black-hole horizon. The line-element (\ref{metric}) employed
in the previous section has failed to accomplish this task. While retaining the assumptions
of staticity and spherical symmetry, we could consider alternative ansatzes for the
line-element of the spacetime that could perhaps allow for more general families
of solutions. To this end, we consider the following line-element
\begin{equation}\label{metric1}
{ds}^2=-e^{A(r)}{dt}^2+e^{B(r)}{dr}^2+H^2(r)({d\theta}^2+\sin^2\theta\,d\varphi^2).
\end{equation}
Note that, by setting $H(r)=r$, we recover the line-element employed in the previous section.
A line-element of this general form was used in the context of the Einstein-scalar-GB theory
in order to find traversable wormhole solutions \cite{ABKKK} -- there, it was demonstrated
that the above line-element led to more extended families of solutions even in cases where
the line-element (\ref{metric}) did not allow for any solutions at all. It seems therefore
justified that such a line-element could be used in our quest for black-hole solutions, too. 

Employing then the line-element (\ref{metric1}), we find the following explicit expressions
for the components of the gravitational field equations
\begin{align}
T^t_{\;\,t}=&\frac{e^{-2B}}{4H^2}\bigg[8 \dot{f} \left(B' \phi ' \left(e^B-3  H'^2\right)-2 e^B
   \phi ''+2  H'^2 \phi ''+4 H' H'' \phi'\right) \nn\\[2mm]&
   + \phi '^2 \left(-16 \ddot{f}
   \left(e^B- H'^2\right)-e^B H^2\right)\bigg]=0\,,\label{Ttt-gen}\\[2mm]
T^r_{\;\,r}=&\frac{e^{-2B}\phi'}{4H^2}\left[e^B H^2 \phi '-8 \dot{f} A' \left(e^B-3  H'^2\right)\right]
=0\,,\label{Trr-gen}\\[2mm]
T^{\theta}_{\;\,\theta}=T^{\varphi}_{\;\,\varphi}=&\frac{e^{-2B}}{4 H}\bigg[+4 \dot{f}
   \left(2 A'' H' \phi '+A' \left(-3 B' H' \phi '+2 H'' \phi '+2 H'
   \phi ''\right)+\left(A'\right)^2 H' \phi '\right)\nn\\[2mm]
   & +\phi '^2 \left(8 A' \ddot{f} H'-e^B H\right)\bigg] =0\,,
\end{align}
and the equation for the scalar field
\begin{align}\label{phi-eq-gen}
&\frac{4\dot{f}e^{-B}}{H}\left[-4 \dot{f} \left(2 A'' \left(e^B- H'^2\right)-A' \left(B'
   \left(e^B-3  H'^2\right)+4 H' H''\right)+ A'^2 \left(e^B- H'^2\right)\right)\right]\nn\\[2mm]
&+  2 H \phi '' + 4 H' \phi '+ H A' \phi '-H B' \phi '=0.
\end{align}

We may, as before, work first with the limit $e^A \rightarrow 0$, or $A' \rightarrow \infty$,
as $r \rightarrow r_h$. Then, Eq. (\ref{Trr-gen}) may be solved to yield an expression
for the dependent function $e^B$: 
\begin{equation}\label{B-fun-gen}
e^{B}=\frac{24 \dot{f} A' H'^2}{8 \dot{f} A'-H^2 \phi '}.
\end{equation}
From the above, we may again find an expression for $B$ that, together with the one
for $e^B$, could be used to eliminate the metric function $B$ from the remaining field
equations. These, then, reduce to a set of three, coupled, second-order, ordinary differential
equations for the metric functions $A$ and $H$ and the scalar field $\phi$, that may
be written as \footnote{This holds under the assumption that $H'' \neq 0$, which corresponds
to the case $H(r)=r$ studied in the previous subsections.}
\beq
A''=\frac{\tilde{P}}{4 \dot f \tilde{S}}\,, \qquad 
\phi''=\frac{\phi'\tilde{Q}}{\tilde{S}}\,, \qquad 
H''=\frac{\tilde{K}}{4 \dot f\tilde{S}}\,. \label{AHphi-sys}
\eeq
In order to reduce the technicalities of our analysis, we present the expression for
$B'$ and the explicit forms of the functions $\tilde S$, $\tilde P$, $\tilde Q$ and $\tilde K$
in Appendix \ref{general-system}. 

Let us instead focus on the physical implications of this system of equations. In the limit
$A'\rightarrow \infty$ while $H$, $\phi$ and $\phi'$ remain finite, we obtain 
\begin{equation}\label{bg5}
e^B=3H'^2+\mathcal{O}\left(\frac{1}{A'}\right).
\end{equation}
By replacing the above in Eqs. (\ref{AHphi-sys}), we get near the point of interest, $r \simeq r_h$,
the results
\bea
&& \hspace*{-0.4cm}\phi''=\;\frac{1}{8}\,\phi'A'+\mathcal{O}(1)\,,\\
&& \hspace*{-0.4cm}A''=-\frac{3}{8}\,A'^2+\mathcal{O}\left(A'\right)\,,\\
&& \hspace*{-0.4cm}H''=\frac{1}{8}\,H'A'+\mathcal{O}(1)\,.\label{bg6}
\eea
Again, in order to keep $\phi''$ finite we demand that $\phi'(r_h)=0$. Integrating twice the
last two equations with respect to the radial coordinate, we obtain the asymptotic solutions
\beq
e^A=a_1 \,(r-r_h)^{8/3} +....\,, \qquad H=h_1\,(r-r_h)^{4/3} + ...\,,
\eeq
where $a_1$ and $h_1$ are integration constants. We observe that the metric function
$e^A$ vanishes indeed at $r=r_h$, as expected near a black-hole horizon. However, the
circumferential radius $H^2(r)$ also vanishes at the same point thus signalling the
presence of an additional pathology in the coordinate system. If we define a new
radial coordinate as $l \equiv {h_1}\,(r-r_h)^{4/3}$, then the line-element (\ref{metric1})
becomes
\beq
ds^2 = -e^{\tilde A}\,dt^2 + e^{\tilde B}\,dl^2 + l^2\,(d\theta^2 +\sin^2\theta\,d\varphi^2)\,,
\eeq
with $e^{\tilde A} \sim a_2\,l^2 +...$ and $e^{\tilde B} \sim b_0=const.$. We may thus conclude that
the obtained asymptotic solution is again not a black hole but more likely a particle-like solution
with similar characteristics as the ones derived in subsection 3.1.

Let us now use the alternative condition that, as $r \rightarrow r_h$, we have
$e^B \rightarrow \infty$, or equivalently $B' \rightarrow \infty$. As in subsection 3.2,
we may then solve Eq. (\ref{Trr-gen}) with respect to $A'$, obtaining the result
\begin{equation}\label{bk1}
A'=\frac{H^2 e^{B} \phi '}{8 \left(e^{B}-3H'^2\right) \dot{f}}\,.
\end{equation}
For a regular black-hole spacetime, we will demand that both the scalar field $\phi$
and the circumferential radius $H$ remain finite near the horizon. Then, there are two
distinct cases:

\begin{itemize}
\item{} If $H'$ remains finite near $r_h$, then, in the limit $e^B \rightarrow \infty$,
Eq. (\ref{bk1}) leads to
\beq
A'\simeq \frac{H^2 \phi'}{8 \dot f} + \mathcal{O}\left(e^{-B}\right). 
\eeq
If, as assumed, $H^2$, $\phi$ and $\phi'$ adopt constant, finite values at the black-hole horizon,
then $A \simeq a_0 + a_1\,(r-r_h)+...$ and the metric component $e^A$ adopts a constant, instead
of a vanishing, value at $r \simeq r_h$.

\item{} If $H'$ is allowed to diverge, then a more careful analysis should be performed.
As an indicative example, we may consider the ``conformal'' case $H= r e^{B/2}$, that
was used in \cite{ABKKK}. Then, Eq. (\ref{bk1}) becomes
\beq
A'=-\frac{e^B\,r^2 \phi'}{2\dot f\left(  8+12 r B' +3 r^2 B'^2 \right)}.
\eeq
For the conventional black-hole dependence where 
$e^B \simeq b_1\,(r-r_h)+...$, the aforementioned equation gives $A' \rightarrow 0$,
thus leading to a constant value for the $g_{tt}$ component near $r_h$. For an
alternative dependence, such as the logarithmic given in Eq. (\ref{B-sol}), we obtain instead
that $A' \rightarrow const.$, which however leads again to a constant $g_{tt}$ 
at $r\simeq r_h$.
\end{itemize}
Therefore, even in the context of the more general line-element (\ref{metric1}), the condition
$e^B \rightarrow \infty$ leads to a solution with similar characteristics as the ones that
emerged in subsection 3.2. Overall, the system of field equations fails to admit an
asymptotic black-hole solution with the anticipated behaviour (\ref{metric-cond})
for both the $g_{tt}$ and $g_{rr}$ metric components. 

%
%

\section{\texorpdfstring{Solutions with Horizons when $\L \neq 0$ }{Solutions with Horizons
for non vanishing cosmological constant}}

In this section, we reinstate the cosmological constant $\Lambda$ in order to investigate
its role in the emergence or not of solutions with a horizon in the context of the pure
scalar-Gauss-Bonnet theory. For $\L \neq 0$, Eq. (\ref{Trr}) takes the form of a 
polynomial with respect to the metric function $e^B$, i.e. $\a e^{2B}+\beta e^{B}+\gamma=0$.
This may be solved to yield
\begin{equation}\label{B-sol-L}
e^B=\frac{-\beta\pm\sqrt{\beta^2-4\a\gamma}}{2\a},
\end{equation}
where:
\beq
\a= -r^2\L V\,, \qquad 
\beta=\frac{r^2{\phi'}^2}{4}-2\dot{f}\phi'A'\,, \qquad \label{p-new}
\gamma=6\dot{f}\phi'A'\,.
\eeq
Taking the derivative of the above expression with respect to the radial coordinate, we may
eliminate $B'$ from the field equations replacing it by 
\begin{equation}\label{r}
B'=-\frac{\gamma'+\beta' e^{B}+\a' e^{2B}}{2\a e^{2B}+\beta e^{B}}.
\end{equation}
The field equations then reduce to a system of two coupled, second-order, ordinary differential
equations for $A$ and $\phi$. This system now has the form
\beq
A''=-\frac{P_1}{S_1}\,, \qquad \quad 
\phi''=-\frac{Q_1}{4 S_1}\,\phi'\,,\label{Aphi-sys-L}
\eeq
where
\begin{align}
P_1=\,&e^{B} \left(-17 r^2 A'^2 \phi '^3 \dot f +64 A'^3 \phi '^2 \dot f ^2
    +18 r A' \phi '^3 \dot f \right)+e^{2 B} \big(-8 r^2 A' \phi '^4 \ddot f
    +9 r^2 A'^2 \phi '^3 \dot f \nn\\[2mm]
   &-20 \Lambda  r^2 A'^2 \phi '  \dot f -8 r A' \phi '^3 \dot f -24 A'^3 \phi '^2
   \dot f ^2+72 \Lambda  r A' \phi' \dot f -r^4 A' \phi '^4+2 r^3 \phi '^4\big)\nn\\[2mm]
   &e^{3 B} \big(4 \dot{f} \Lambda  r^2 A'^2 \phi '+6\dot{f} r A'  \phi '^3
   -96 \dot{f} \Lambda  r A' \phi '-2 \Lambda  r^4 A'\phi '^2-r^3
   \phi '^4+4 \Lambda  r^3 \phi'^2\big)\nn\\[2mm]
  &+ e^{4 B} \left(24 \dot{f} \Lambda  r A' \phi '+8 \Lambda ^2 r^4
   A'-16 \Lambda ^2 r^3\right) +16 e^{5 B} \Lambda ^2 r^3-72 \dot{f}^2 A'^3 \phi '^2,
  \end{align}
\begin{align}
Q&=e^B \big(36 \dot{f} r^2 A' \phi '^3+896 \dot{f} A' \ddot{f} \phi '^3-192 \dot{f}^2 A'^2
   \phi '^2-16 r^2 \ddot{f}\phi'^4-120 \dot{f} r \phi '^3\big)\nn\\[2mm]
   &+e^{2 B} \big(-20 \dot{f} r^2 A'\phi '^3+144 \dot{f}\Lambda  r^2 A' \phi '
   -320 \dot{f} A' \ddot{f} \phi'^3+32 \dot{f}^2 A'^2 \phi'^2\nn\\[2mm]
   &+16 r^2 \ddot{f} \phi '^4+128 \Lambda  r^2
   \ddot{f} \phi '^2+64 \dot{f} r \phi '^3-96
   \dot{f} \Lambda  r \phi '+r^4 \phi '^4\big)\nn\\[2mm]
   &+e^{3 B} \big(-80 \dot{f} \Lambda  r^2 A' \phi '-128 \Lambda  r^2
   \ddot{f} \phi '^2-8 \dot{f} r \phi '^3+128
   \dot{f} \Lambda  r \phi '-4 \Lambda  r^4 \phi'^2\big)\nn\\[2mm]
   &+e^{4 B} \big(-32 \dot{f}\Lambda  r \phi '-32 \Lambda ^2 r^4\big)
   -576 \dot{f} A' \ddot{f} \phi '^3+288 \dot{f}^2 A'^2 \phi '^2,
\end{align}
\begin{align}
S_1&=8 \dot{f} \phi ' \left[ 2 \dot{f} A' \phi ' \left(-14 e^B+5 e^{2 B}+9\right)
    -e^B \left(e^B-2\right) r^2 \phi '^2+4
   e^{2 B} \left(e^B-1\right) \Lambda  r^2\right].
\end{align}

We will assume that, near the black-hole horizon $r_h$, $e^A \rightarrow \infty$, or
equivalently $A' \rightarrow \infty$. Then, Eq. (\ref{B-sol-L}) takes the expanded form:
\begin{align}\label{expb}
e^B=\frac{-2 \dot f \phi '}{ r^2 \Lambda  
   V} A'  + \frac{-24\L V \dot f \f'+2 \dot f \f'^3}{8 \L V \dot f \f'} + \mathcal{O}\left(\frac{1}{A'}\right).
\end{align}
If we replace the above expansion, too, into the system (\ref{Aphi-sys-L}), we obtain the
following equations in the limit $A' \rightarrow \infty$:
\begin{align}
A''=&-\frac{2 \dot f \f'}{r^3 \L V}\,A'^2+\mathcal{O}\left(A'\right),\label{expa}\\[1mm]
\f''=&\left(1-\frac{2 \dot f \f'}{r^3 \L V}\right) \phi'A' + \mathcal{O}(1).\label{expf}
\end{align}
The finiteness of $\f''$ at the horizon dictates that the following relation should hold
\begin{align}\label{solf}
\f'_h=\frac{r_h^3 \L}{2\dot f (\f_h)}.
\end{align}
The above condition on the first derivative of the scalar field at the horizon is analogous 
to the one derived in the context of the Einstein-scalar-Gauss-Bonnet theory -- there,
that condition also related $\phi'_h$ to the parameters of the theory and ensured the
regularity of the black-hole horizon. Here, we observe that the condition (\ref{solf})
relates $\phi'$  with the value of the cosmological constant. 

Let us proceed to derive the complete form of the asymptotic solution near $r_h$.
By using the above condition on $\phi'_h$, the system (\ref{expa})-(\ref{expf}) takes
the simplified form:
\begin{align}
A''=&-A'^2+\mathcal{O}\left(A'\right),\label{expa2}\\[3mm]
\f''=& \mathcal{O}(1).\label{expf2}
\end{align}
Again, this resembles the behaviour obtained around a black-hole horizon. Integrating
Eq. (\ref{expa2}) with respect to $r$, we obtain the usual behaviour
\beq
A'=\frac{1}{r-r_h}\,,\label{ahorp}
\eeq
while a second integration leads to the complete asymptotic solution for the metric
functions and the scalar field near $r_h$
\begin{align}
e^A&=a_1\,(r-r_h)+... , \label{ahor}\\[2mm]
e^{-B}&=b_1\,(r-r_h)+... , \label{bhor}\\[2mm]
\f&=\f_h+\f_h'(r-r_h)+\f_h''(r-r_h)^2+...\,. \label{fhor}
\end{align}
The above describes indeed a solution with a horizon and a regular scalar field. There is,
however, a caveat: if we replace $\phi'_h$ from Eq. (\ref{solf}) into Eq. (\ref{expb}), we find: 
\begin{equation}
b_1=-1/r_h.
\end{equation}
The robustness of the metric is then ensured only if Eqs. (\ref{ahor})-(\ref{fhor}) are
re-written as
\begin{align}
e^A&=|a_1|\,(r_c-r)+... ,\label{a-cosm}\\[2mm]
e^{-B}&=\frac{1}{r_c}\,(r_c-r)+...\\[2mm]
\f&=\f_c+\f_c'(r_c-r)+\f_c''(r_c-r)^2+...\,.\label{f-cosm}
\end{align}
Note that $a_1$ is a completely arbitrary constant and thus may be appropriately chosen.
The above reveal that, when $\L >0$, the asymptotic solution we have found corresponds to
a cosmological horizon $r_c$ with a non-trivial scalar field. Such solutions have indeed
been derived in the context of the pure scalar-Gauss-Bonnet theory but from the perspective
of a homogeneous, isotropic universe \cite{KGD}. Here, we have in fact re-derived those
solutions describing a de Sitter universe using isotropic, spherically-symmetric coordinates.
In the case of an Anti-de Sitter spacetime with $\Lambda <0$, no cosmological horizon
exists and no robust black-hole solution emerges either in the context of our theory.
Therefore, even in the presence of a non-vanishing cosmological constant, the pure
scalar-Gauss-Bonnet theory fails to support a regular black-hole solution with 
scalar hair.

We close this section by noticing that the asymptotic solution (\ref{a-cosm})-(\ref{f-cosm})
emerges also in the case where we work with the assumption that near the horizon it is
$e^B$, or $B'$, that diverges; this is expected for a physical, spherically-symmetric horizon.
Also, similar solutions which possess again a cosmological
rather than a black-hole horizon emerge if we employ the more general line-element
(\ref{metric1}) - as their characteristics are similar to the ones derived above,
we refrain from presenting the corresponding analysis.


\section{Synergy between the Ricci and GB terms}

In order to shed more light on the fact that the pure scalar-Gauss-Bonnet theory does
not admit black-hole solutions but allows for cosmological solutions with a horizon to emerge,
we turn to the complete Einstein-scalar-GB theory. In the context of the latter, we will
investigate the contribution of each gravitational term to the formation of a black hole.
To this end, we choose the case of the exponential coupling function $f(\phi)=\alpha e^{-\phi}$,
where $\alpha$ is a positive coupling constant. This choice leads to the dilatonic black holes
that were found in \cite{DBH} and studied again in \cite{ABK1}. They are a one-parameter family
of black-hole solutions, with the independent parameter being any one of the three parameters
of the theory, namely $(\alpha, r_h, \phi_h)$, due to the rescaling symmetries of the theory. Here,
we fix the horizon radius $r_h$ to unity while the value of the scalar field at the horizon
$\phi_h$ is determined by the boundary condition $\phi_\infty=1$ at infinity.  
These leave the coupling constant $\alpha$ as the independent parameter of the theory. 

A theoretical argument for the regularity of the formed black-hole horizon, analogous to
the one that led to Eq. (\ref{solf}), leads to the following condition on the first 
derivative of the scalar field at the horizon \cite{ABK1}
\begin{equation}\label{con-phi'}
\phi'_h=\frac{r_h}{4\dot{f}_h}\left(-1\pm\sqrt{1-\frac{96\dot{f}_h^2}{r_h^4}}\right).
\end{equation}
The reality of the value of $\phi'_h$ imposes the following additional constraint on
the coupling function
\begin{equation}\label{fcon}
\dot{f}_h^2<\frac{r_h^4}{96}\,.
\end{equation}
For the choice $f(\phi)=\alpha e^{-\phi}$, the above reduces to a constraint on the maximum
allowed value of the coupling parameter, that is 
\beq 
\frac{\alpha}{r_h^2} \simeq 0.123\,.
\label{a-bound}
\eeq

\begin{table}[t!]
\begin{center}
$\begin{array}{|c|c|c|} \hline\hline & & \\[-0.35cm] 
\hspace*{1cm}  \alpha/r_h^2 \hspace*{1cm}&  \hspace*{1.5cm}
R \hspace*{1.5cm} & \hspace*{0.5cm} f(\phi) R^2_{GB} \hspace*{1.0cm} \\[0.05cm] \hline \hline
& & \\[-0.35cm]
0.001 & 3.318 \times 10^{-9} & 0.004 \\ 
0.006 & 5.002 \times 10^{-8} & 0.025 \\ 
0.015 &  1.536 \times 10^{-6} & 0.070 \\ 
0.025 & 1.112 \times 10^{-5} & 0.117 \\ 
0.037 & 6.134 \times 10^{-5} & 0.182 \\ 
0.051 & 2.835 \times 10^{-4} & 0.270 \\ 
0.065 & 9.790 \times 10^{-4} & 0.372 \\ 
0.075 & 2.051 \times 10^{-3} & 0.452 \\ 
0.087 & 4.897 \times 10^{-3} & 0.571 \\ 
0.097 & 9.722 \times 10^{-3} & 0.691 \\ 
0.106 & 1.962 \times 10^{-2} & 0.847 \\ 
0.116 & 4.120 \times 10^{-2} & 1.067 \\ 
0.123 & 8.097 \times 10^{-2} & 1.320 \\ [0.05cm] 
%
\hline \end{array}$
\end{center}
\caption{An indicative list of values of the coupling constant $\alpha$ and the values of
the Ricci scalar $R$ and the combination $f(\phi) R^2_{GB}$ near the black-hole horizon $r_h$.
\label{R-GB}}
\end{table}

Black-hole solutions then arise in the context of the Einstein-scalar-GB theory for the range
of values $[0, 0.123]$ of the coupling constant $\alpha$. When $\alpha \rightarrow 0$,
the GB term decouples from the theory and the only black-hole solution is the Schwarzschild
solution characterized by a trivial scalar field. As $\alpha$ adopts a non-vanishing value,
a GB black hole is formed that possess scalar hair and a horizon radius smaller than 
the one of the Schwarzschild solution \cite{ABK1}. In Table 1, we display a number of indicative
values of $\alpha$ in the range $[0, 0.123]$ and the values of the Ricci scalar and the
combination $f(\phi)\,R^2_{GB}$ near the horizon for each of the corresponding black-hole
solution. We observe that for small values of the coupling constant, the Ricci term adopts
a very small value - this is due to the fact that, for small $\alpha$, the obtained solution
is still very close to the Schwarzschild one that is a vacuum solution with $R=0$. The
GB term, in contrast, adopts a much larger value from the beginning due to the contribution
of the Riemann tensor that is not zero even for the Schwarzschild solution. As $\alpha$
moves towards its maximum value, the curvature of spacetime increases and this enhances
the magnitude of both gravitational terms. Above the maximum value $a_{max}=0.123$,
no black-hole solutions emerge. 

We also observe that the sign of $R$ remains always positive, which reflects the positive
curvature of spacetime around the formed horizon. In the Einstein-Hilbert action, the presence
of the Ricci term leads to the attractive force of gravity. The presence of the GB term, on the
other hand, produces the opposite effect as it leads to a repulsive force in the theory.
Indeed, if we examine the components of the effective energy-momentum tensor near
the horizon, we obtain the following expressions for the effective energy-density and radial
pressure of the system:
\begin{align}
\rho=-T^t_{\;\,t}=&-\frac{2e^{-B}}{r^2}B'\phi'\dot{f}+\mathcal{O}(r-r_h)\,,\label{rho}\\
p_r=T^r_{\;\,r}=&-\frac{2e^{-B}}{r^2}A'\phi'\dot{f}+\mathcal{O}(r-r_h)\,.\label{p}
\end{align}
Note that the contribution of the scalar-field kinetic term vanishes near the horizon and
the dominant contribution to the energy-momentum tensor components comes from its
effective potential, i.e. by its coupling to the GB term. The combination $\phi'\dot{f}$ is
always negative at the horizon due to the regularity constraint (\ref{con-phi'}).
Also, near the horizon radius, it holds that $A' \simeq -B' \simeq 1/(r-r_h) + ...$, since
$e^A$ increases near the horizon, as $r$ increases, while $e^B$ decreases. These, then,
lead to an effective, local equation of state for the GB contribution of the form
\beq
p_r=-\rho = \frac{2 b_1}{r_h^2}\,\big|(\phi'\dot{f})_h\big|>0\,. \label{eq-state}
\eeq
Clearly, the GB term saturates the dominant energy condition and violates the weak and
strong energy conditions by creating a negative effective energy-density and an equal
in magnitude, but positive, radial pressure component at the horizon. It is this violation
of the energy conditions that causes the evasion of the novel no-hair theorem
\cite{Bekenstein} and allows for the emergence of black holes with non-trivial scalar hair
\cite{DBH, ABK1}. The positive, outward radial pressure signifies in addition the repulsive
role of the GB term in the system.

We may, therefore, interpret now the non-emergence of a black-hole horizon in the
context of the pure scalar-GB theory. The presence of the Ricci term is necessary in
order to provide the attractive force that will create the positively-curved topology
around the formed black hole. Even in vacuum, the Ricci term causes the formation
of a black hole in the form of the Schwarzschild solution. When the GB term is turned
on, the Schwarzschild solution gets {\it naturally} scalarised \footnote{As opposed
to {\it spontaneous} scalarisation, that takes place when a tachyonic scalar mode
causes the system to shift from the unstable Schwarzschild solution to a more stable
solution with a non-trivial scalar field. This scalarisation is realised only for a
particular regime of the independent parameter where the Schwarzschild solution
is destabilised.} as it is automatically donned with a non-trivial, regular scalar field.
Due to the repulsive effect of the GB term, the black-hole horizon is now formed 
at a smaller horizon value than in the Schwarzschild case -- indeed, in \cite{ABK1},
it was demonstrated that, for a fixed coupling constant, a GB black hole has always
a smaller horizon radius compared to the one of the Schwarzschild solution with the
same mass. Apparently, gravity dominates over a smaller regime of spacetime, 
creating a black-hole topology, in the presence of the GB term. In other words, the
same amount of mass needs to be ``squeezed" more to create a black hole when
the repulsive GB term is present in the theory. 

Therefore, the GB term makes the formation of a black hole more difficult. What it does
facilitate is the dressing of the black-hole solution with a non-trivial scalar field, a
feature that would have been forbidden in its absence. As the value of the coupling
constant increases, the weight of the GB term in the theory gradually increases, too.
The same holds for its repulsive effect. Beyond the maximum value (\ref{a-bound})
of the GB coupling parameter, no black-hole horizon can be formed -- or sustained --
any more. We may easily then justify the fact that a pure scalar-GB theory can not, in the
absence of the Ricci scalar, create by itself a black-hole solution.
 
In the case of cosmological solutions, the presence of the GB term in the theory
leads to the emergence of singularity-free \cite{ART, KRT, KGD} and inflationary
solutions \cite{KGD}. In a cosmological context, traditional gravity, in the form of
the Ricci term, leads to the formation of the initial singularity, a feature that is
not desirable in the theory. The addition of the GB term with its repulsive effect
manages to provide the necessary outward pressure to the system so that the
initial singularity is avoided, and a smooth transition between a collapsing and
an expanding phase of the Universe is realised \cite{ART, KRT, KGD}. In the
emergence of de Sitter, inflationary solutions \cite{KGD}, the GB term is again
providing the outward pressure component that accelerates the expansion of
the Universe even in the absence of any potential for the scalar field.   

%
%
\section{Conclusions}

After the derivation, in an analytical way, of cosmological singularity-free and inflationary
solutions in the context of the pure scalar-Gauss-Bonnet theory \cite{KGD}, here, we have 
investigated whether black-hole solutions can be supported in the context of the
same theory. We have therefore ignored the presence of all terms associated with
the Ricci term in the field equations and used both analytical and numerical means
to integrate them. Any solutions that could emerge would rely solely on the
synergy between the scalar field and the GB term in the theory and would 
correspond to zero-energy and zero-pressure gravitational solutions. 

We initially focused on the derivation of static, spherically-symmetric solutions that
would describe a regular black hole with a non-trivial scalar hair. In section 3, we
solved analytically
the field equations near the sought-for black-hole horizon as it is only there that
ignoring the Ricci scalar, compared to the quadratic GB term, may be justified.
Working with the assumption that there the $g_{tt}$ metric component vanishes,
we derived a family of gravitational solutions with a finite $g_{rr}$
and a regular scalar field. The spacetime possesses a true singularity but a 
finite energy-momentum tensor, a profile that resembles the one of particle-like
solutions in quadratic gravitational theories \cite{Afonso, KKK2}. Alternatively,
demanding that the $g_{rr}$ diverges at a specific value of the radial coordinate,
we determined, first analytically and then numerically, a second family of
gravitational solutions with no spacetime singularity and a finite, again, 
energy-momentum tensor. We have postponed the study of these two families
of solutions and their physical interpretation for a future study.

Coming back to our quest for black-hole solutions, in Section 4, we considered a
more general line-element for the spacetime, that preserved the assumptions of
staticity and spherical-symmetry. Despite the increased flexibility of the line-element,
that resulted in the addition of a third unknown metric function, no solutions with
a black-hole horizon were found. The asymptotic solutions derived analytically
 in the small-$r$ regime shared the same characteristics as the ones found in
Section 3. 

Re-instating the cosmological constant, that was ignored in the first part of our
analysis, we looked again for solutions with a horizon. The analysis in this case
resembled the one that led to the derivation of regular black-hole solutions with
scalar hair \cite{ABK1}. Indeed, solutions with a horizon and a regular scalar
field were successfully found in Section 5 in the case of a positive cosmological
constant, however, these were shown to correspond to a cosmological rather than
to a black-hole horizon. 

In Section 6, we looked more carefully at the synergy between the Ricci and GB terms.
The values of the components of the energy-momentum tensor near the black-hole
horizon reveal the repulsive effect of the GB term as opposed to the attractive effect
of the Ricci term. As the latter is necessary in order to create the positive curvature
around a black hole, no such solution emerges in the presence of only the GB term,
that in fact works towards pushing outwards any distribution of matter. Even in the
context of the complete Einstein-scalar-GB theory, black holes emerge only up to
a maximum value of the GB coupling parameter -- it is only over this restricted
parameter regime that the Ricci term manages to form a black-hole horizon despite
the presence of the GB term. Nevertheless, the GB term justifies its presence by
supporting a non-trivial, regular scalar field, a feature forbidden by General
Relativity. 

In the context of the effective-field-theory point of view, one could imagine of 
adding even higher-derivative terms in the quadratic action (\ref{action}). How 
are then the previously-derived solutions \cite{ABK1, BAK} modified by the
presence of these gravitational corrections? In the light of the analysis of Section 6,
we conclude that this depends on the role of these terms when it comes to their
contributions to the effective energy-momentum tensor. If such a higher-derivative
term had a positive contribution to the radial pressure as the GB term, then
it would have a destabilising effect and black-hole solutions could emerge
for a more restricted range of values for the coupling constant $\alpha$. If,
on the other hand, the produced contribution to the radial pressure of the system
was negative, the added term would have a role similar to that of the Ricci scalar
and the emergence of  black-hole solutions would be facilitated.

Closing this work, let us return to the question we posed in Section 2 to which we may
now give a simple answer: we may indeed find spacetime regimes where the GB term is
overwhelmingly dominant over the Ricci scalar, however, the gravitational solution
that would form will not be a black hole. Nevertheless, solutions the existence of which
does not rely on the attractive nature of the Ricci term, and thus on its presence in the
theory, do exist and these include a plethora of interesting solutions such as particle-like
solutions, cosmological solutions and even wormholes. As that part of the phase-space
of solutions of the pure scalar-GB theory has not adequately explored so far, we plan
to return to such a careful study soon.


\section*{Acknowledgments}

We would like to thank Naresh Dadhich for valuable discussions during the early stages
of this work. This research is implemented through the Operational Program ``Human Resources
Development, Education and Lifelong Learning" and is co-financed by the European
Union (European Social Fund) and Greek national funds.

\appendix

\numberwithin{equation}{section}	

\section{Scalar Quantities}
\label{scalar}

By employing the metric components of the line-element (\ref{metric}), one may
compute the following scalar-invariant gravitational quantities:
\bea
R&=&+\frac{e^{-B}}{2r^2}\left(4e^B-4-r^2A'^2+4rB'-4rA'+r^2A'B'-2r^2A''\right),\label{A1}\\\nonumber\\
R_{\mu\nu}R^{\mu\nu}&=&+\frac{e^{-2B}}{16 r^4}\left[8(2-2e^B+rA'-rB')^2+r^2(rA'^2-4B'-rA'B'+2rA'')^2\right.\nonumber\\
&&\left.+r^2(rA'^2+A'(4-rB')+2rA'')^2\right],\\\nonumber\\
R_{\mu\nu\rho\sigma}R^{\mu\nu\rho\sigma}&=&+\frac{e^{-2B}}{4r^4}\left[r^4A'^4-2r^4A'^3B'-4r^4A'B'A''+r^2A'^2(8+r^2B'^2+4r^2A'')\right.\nonumber\\
&&\left.+16(e^B-1)^2+8r^2B'^2+4r^4A''^2\right],\\\nonumber\\
R_{GB}^2&=&+\frac{2e^{-2B}}{r^2}\left[(e^B-3)A'B'-(e^B-1)A'^2-2(e^B-1)A''\right].\label{A4}
   \eea


\section{General System of Coupled Equations}
\label{general-system}

Employing the expression (\ref{B-fun-gen}) for the metric function $e^B$, we may easily
compute the first derivative of the metric function $B(r)$. This is found to have the form
\begin{align}
B'=&\frac{e^{-B}}{r \phi ' \big(H^2 \phi '-8 \dot{f} A'\big)}\bigg\{ -2 \phi ' \left[\phi ' \left(e^B H \left(r H'+H\right)-4 r A' \ddot{f}
   \left(e^B-3H'^2\right)\right)+e^B H^2 r \phi ''\right]\nn \\[2mm]  
   &+ 8 \dot{f} \Bigl[r A'' \phi ' \left(e^B-3  H'^2\right)+
   r A' \phi '' \left(e^B-3 H'^2\right)+2 A'\phi '
   \left(e^B-3 H' \left(r H''+H'\right)\right)\Bigr]  \bigg\}.\label{bg2}
\end{align}
Then, the field equations (\ref{Ttt-gen})-(\ref{phi-eq-gen}) reduce to a set of three, coupled,
second-order, ordinary differential equations for the metric functions $A$ and $H$ and the
scalar field $\phi$, given in Eqs. (\ref{AHphi-sys}). The functions $\tilde S$, $\tilde P$,
$\tilde Q$ and $\tilde K$ have now the following explicit forms
\begin{align}
\tilde{S}=&8 r \left[2 \dot{f} A' \left(e^B-H'^2\right) \left(e^B+3
   H'^2\right)+e^B H^2 H'^2 \phi '\right],\label{tilde-S} \\\nonumber\\
\tilde{P}=&8 \dot{f}^2 A'^2 \left(3 r A'+4\right) \left(e^B-H'^2\right) \left(e^B+3
   H'^2\right)+e^B H^2 r  \phi '^2 \Big[-4 e^B H H'\nn\\[2mm]
   &A' \left(16 \ddot{f} \left(e^B+H'^2\right)+e^B H^2\right)\Big]
   -4 e^B \dot{f} H A' \phi ' \Big[2 e^B r \left(H A'+6 H'\right)\nn\\[2mm]
   &+ H'^2 \left(H r A'+4 r H'-4 H\right)\Big],\\\nonumber\\
\tilde{Q}=&\phi ' \Big[-4 e^B H'^2 \left(r A' \left(8
   \ddot{f}+H^2\right)+H \left(5 r H'+H\right)\right)-e^{2 B} r
   \left(A' \left(16 \ddot{f}+H^2\right)-4 H H'\right)\nn\\[2mm]
   &+48 r A' \ddot{f}
   H'^4\Big]+2 \dot{f} A' \left(r A'-4\right) \left(e^B-H'^2\right)
   \left(e^B+3 H'^2\right), \\\nonumber\\
\tilde{K}=& 8\dot{f}^2 A' H' \left(r A'-4\right) \left(e^B-H'^2\right) \left(e^B+3
   H'^2\right)+ e^B H^2 r H' \phi '^2 \left[16 \ddot{f}
   \left(e^B-H'^2\right)+e^B H^2\right]\nn\\[2mm]
   &+4 e^B \dot{f} H \phi ' \left[-e^B r H' \left(H A'+12 H'\right)+2
   H'^3 \left(H r A'+5 r H'-2 H\right)+2 e^{2 B} r\right]. \label{tilde-K}
\end{align}




\begin{thebibliography}{9}

\bibitem{LIGO}
https://www.ligo.org/

\bibitem{VIRGO}
http://www.virgo-gw.eu/

\bibitem{Stelle} K.~S.~Stelle,
  Phys.\ Rev.\ D {\bf 16} (1977) 953.

\bibitem{General} T.~P.~Sotiriou,
  Lect.\ Notes Phys.\  {\bf 892} (2015) 3;\\
E.~Berti {\it et al.},
  Class.\ Quant.\ Grav.\  {\bf 32} (2015) 243001.

 
\bibitem{NH-scalar} J.~D.~Bekenstein,
  Phys.\ Rev.\ Lett.\  {\bf 28} (1972) 452; 
C.~Teitelboim,
  Lett.\ Nuovo Cim.\  {\bf 3S2} (1972) 397.

\bibitem{YM} M.~S.~Volkov and D.~V.~Galtsov,
 JETP Lett.\  {\bf 50} (1989) 346;
P.~Bizon,
  Phys.\ Rev.\ Lett.\  {\bf 64} (1990) 2844;
B.~R.~Greene, S.~D.~Mathur and C.~M.~O'Neill,
  Phys.\ Rev.\ D {\bf 47} (1993) 2242;
K.~i.~Maeda, T.~Tachizawa, T.~Torii and T.~Maki,
Phys.\ Rev.\ Lett.\  {\bf 72} (1994) 450.

\bibitem{Skyrmions} H.~Luckock and I.~Moss,
  Phys.\ Lett.\ B {\bf 176} (1986) 341;
S.~Droz, M.~Heusler and N.~Straumann,
  Phys.\ Lett.\ B {\bf 268} (1991) 371.

\bibitem{Conformal} J.~D.~Bekenstein,
  Annals Phys.\  {\bf 82} (1974) 535; 
Annals Phys.\  {\bf 91} (1975) 75.

\bibitem{Bekenstein} J.~D.~Bekenstein,
  Phys.\ Rev.\ D {\bf 51} (1995) no.12,  R6608.

\bibitem{Zwiebach}
  B.~Zwiebach,
  Phys.\ Lett.\  {\bf 156B}, 315 (1985).

\bibitem{Gross}
  D.~J.~Gross and J.~H.~Sloan,
  Nucl.\ Phys.\ B {\bf 291}, 41 (1987).

\bibitem{Metsaev}
  R.~R.~Metsaev, A.~A.~Tseytlin,
  Nucl.\ Phys.\  {\bf B293 }, 385 (1987).

\bibitem{DBH} P.~Kanti, N.~E.~Mavromatos, J.~Rizos, K.~Tamvakis and E.~Winstanley,
  Phys.\ Rev.\ D {\bf 54} (1996) 5049; 
Phys.\ Rev.\ D {\bf 57} (1998) 6255.
  
\bibitem{Gibbons}
  G.~W.~Gibbons and K.~i.~Maeda,
  Nucl.\ Phys.\ B {\bf 298} (1988) 741.
  
\bibitem{Callan}
  C.~G.~Callan, Jr., R.~C.~Myers and M.~J.~Perry,
  Nucl.\ Phys.\ B {\bf 311} (1989) 673.
  
\bibitem{Campbell}
  B.~A.~Campbell, M.~J.~Duncan, N.~Kaloper and K.~A.~Olive,
  Phys.\ Lett.\ B {\bf 251} (1990) 34;
 B.~A.~Campbell, N.~Kaloper and K.~A.~Olive,
  Phys.\ Lett.\ B {\bf 263} (1991) 364.

\bibitem{Mignemi}
S.~Mignemi and N.~R.~Stewart,
  Phys.\ Rev.\ D {\bf 47} (1993) 5259.
  
\bibitem{Kanti1995}
  P.~Kanti and K.~Tamvakis,
  Phys.\ Rev.\ D {\bf 52} (1995) 3506.

\bibitem{Torii}
  T.~Torii, H.~Yajima and K.~i.~Maeda,
  Phys.\ Rev.\ D {\bf 55} (1997) 739.
  
\bibitem{KT}
  P.~Kanti and K.~Tamvakis,
  Phys.\ Lett.\ B {\bf 392} (1997) 30; 
P.~Kanti and E.~Winstanley,
  Phys.\ Rev.\ D {\bf 61} (2000) 084032.

\bibitem{Guo} Z.~K.~Guo, N.~Ohta and T.~Torii,
 Prog.\ Theor.\ Phys.\  {\bf 120} (2008) 581;
 K.~i.~Maeda, N.~Ohta and Y.~Sasagawa,
  Phys.\ Rev.\ D {\bf 80} (2009) 104032;
N.~Ohta and T.~Torii,
  Prog.\ Theor.\ Phys.\  {\bf 124} (2010) 207.
  
 \bibitem{Kleihaus}
  B.~Kleihaus, J.~Kunz and E.~Radu,
  Phys.\ Rev.\ Lett.\  {\bf 106} (2011) 151104;
  B.~Kleihaus, J.~Kunz, S.~Mojica and E.~Radu,
  Phys.\ Rev.\ D {\bf 93} (2016) no.4,  044047.
  
   \bibitem{Pani}
  P.~Pani, C.~F.~B.~Macedo, L.~C.~B.~Crispino and V.~Cardoso,
  Phys.\ Rev.\ D {\bf 84} (2011) 087501;
  P.~Pani, E.~Berti, V.~Cardoso and J.~Read,
  Phys.\ Rev.\ D {\bf 84} (2011) 104035.
  
\bibitem{Herdeiro}
  C.~A.~R.~Herdeiro and E.~Radu,
  Phys.\ Rev.\ Lett.\  {\bf 112} (2014) 221101.
  
  \bibitem{Ayzenberg}
  D.~Ayzenberg and N.~Yunes,
  Phys.\ Rev.\ D {\bf 90} (2014) 044066
   Erratum: [Phys.\ Rev.\ D {\bf 91} (2015) no.6,  069905].

\bibitem{Win-review} E.~Winstanley,
  Lect.\ Notes Phys.\  {\bf 769} (2009) 49.

  \bibitem{Charmousis-rev}
  C.~Charmousis,
  Lect.\ Notes Phys.\  {\bf 769} (2009) 299.
  
 \bibitem{Herdeiro-review}
  C.~A.~R.~Herdeiro and E.~Radu,
  Int.\ J.\ Mod.\ Phys.\ D {\bf 24} (2015) no.09,  1542014.
  
   \bibitem{Blazquez}
  J.~L.~Blazquez-Salcedo {\it et al.},
  IAU Symp.\  {\bf 324} (2017) 265.

\bibitem{Horndeski} G.~W.~Horndeski,
  Int.\ J.\ Theor.\ Phys.\  {\bf 10} (1974) 363.

\bibitem{Galileon}
  A.~Nicolis, R.~Rattazzi and E.~Trincherini,
  Phys.\ Rev.\ D {\bf 79} (2009) 064036.
  
\bibitem{SF} T.~P.~Sotiriou and V.~Faraoni,
  Phys.\ Rev.\ Lett.\  {\bf 108} (2012) 081103.

\bibitem{HN}  L.~Hui and A.~Nicolis,
  Phys.\ Rev.\ Lett.\  {\bf 110} (2013) 241104.
  
\bibitem{SZ}  T.~P.~Sotiriou and S.~Y.~Zhou,
  Phys.\ Rev.\ Lett.\  {\bf 112} (2014) 251102.
  
\bibitem{Babichev}
  E.~Babichev and C.~Charmousis,
  JHEP {\bf 1408} (2014) 106.
   
\bibitem{Benkel}  T.~P.~Sotiriou and S.~Y.~Zhou,
  Phys.\ Rev.\ D {\bf 90} (2014) 124063;
R.~Benkel, T.~P.~Sotiriou and H.~Witek,
  Class.\ Quant.\ Grav.\  {\bf 34} (2017) no.6,  064001;
Phys.\ Rev.\ D {\bf 94} (2016) no.12,  121503.

\bibitem{Yunes2011}
  N.~Yunes and L.~C.~Stein,
  Phys.\ Rev.\ D {\bf 83} (2011) 104002.
    
\bibitem{ABK1} G. Antoniou, A. Bakopoulos and P. Kanti, 
  Phys.\ Rev.\ Lett.\  {\bf 120} (2018) no.13,  131102;
  Phys.\ Rev.\ D {\bf 97} (2018) no.8,  084037.

\bibitem{Doneva} D.~D.~Doneva and S.~S.~Yazadjiev,
  Phys.\ Rev.\ Lett.\  {\bf 120} (2018) no.13,  131103.

\bibitem{Silva} H.~O.~Silva, J.~Sakstein, L.~Gualtieri, T.~P.~Sotiriou and E.~Berti,
  Phys.\ Rev.\ Lett.\  {\bf 120} (2018) no.13,  131104.

\bibitem{Bardoux}
  Y.~Bardoux, M.~M.~Caldarelli and C.~Charmousis,
  JHEP {\bf 1205} (2012) 054.

\bibitem{Ayzen} K.~Yagi, L.~C.~Stein, N.~Yunes and T.~Tanaka,
 Phys.\ Rev.\ D {\bf 85} (2012) 064022
 Erratum: [Phys.\ Rev.\ D {\bf 93} (2016) no.2,  029902];
  D.~Ayzenberg, K.~Yagi and N.~Yunes,
  Phys.\ Rev.\ D {\bf 89} (2014) no.4,  044023.

\bibitem{Charmousis}
  C.~Charmousis, T.~Kolyvaris, E.~Papantonopoulos and M.~Tsoukalas,
  JHEP {\bf 1407} (2014) 085.

\bibitem{Correa}
  F.~Correa, M.~Hassaine and J.~Oliva,
  Phys.\ Rev.\ D {\bf 89} (2014) no.12,  124005.

\bibitem{Dolan}
  S.~R.~Dolan, S.~Ponglertsakul and E.~Winstanley,
  Phys.\ Rev.\ D {\bf 92} (2015) no.12,  124047.

\bibitem{Kunz} 
J.~L.~Blazquez-Salcedo, C.~F.~B.~Macedo, V.~Cardoso, V.~Ferrari, L.~Gualtieri, F.~S.~Khoo, J.~Kunz and P.~Pani,
Phys.\ Rev.\ D {\bf 94} (2016) no.10,  104024.

\bibitem{Bhatta} S.~Bhattacharya and S.~Chakraborty,
  Phys.\ Rev.\ D {\bf 95} (2017) no.4,  044037;
  I.~Banerjee, S.~Chakraborty and S.~SenGupta,
  Phys.\ Rev.\ D {\bf 96} (2017) no.8,  084035.

\bibitem{Doneva-NS} D.~D.~Doneva and S.~S.~Yazadjiev,
  JCAP {\bf 1804} (2018) no.04,  011.

\bibitem{Motohashi} H.~Motohashi and M.~Minamitsuji,
  Phys.\ Lett.\ B {\bf 781} (2018) 728;
  Phys.\ Rev.\ D {\bf 98} (2018) no.8,  084027.

\bibitem{Radu} C.~A.~R.~Herdeiro, E.~Radu, N.~Sanchis-Gual and J.~A.~Font,
  Phys.\ Rev.\ Lett.\  {\bf 121} (2018) no.10,  101102;
T.~Delsate, C.~Herdeiro and E.~Radu,
  Phys.\ Lett.\ B {\bf 787} (2018) 8;
  Y.~Brihaye, C.~Herdeiro and E.~Radu,
  Phys.\ Lett.\ B {\bf 788} (2019) 295.

\bibitem{Doneva-Papa}
  D.~D.~Doneva, S.~Kiorpelidi, P.~G.~Nedkova, E.~Papantonopoulos and S.~S.~Yazadjiev,
  Phys.\ Rev.\ D {\bf 98} (2018) no.10,  104056.

\bibitem{Butler} M.~Butler, A.~M.~Ghezelbash, E.~Massaeli and M.~Motaharfar,
  arXiv:1808.03217 [hep-th].

\bibitem{Danila}
  B.~Danila, T.~Harko, F.~S.~N.~Lobo and M.~K.~Mak,
  arXiv:1811.02742 [gr-qc].

\bibitem{Stetsko}
  M.~M.~Stetsko,
  arXiv:1811.05030 [hep-th].

\bibitem{Tatter}
  O.~J.~Tattersall, P.~G.~Ferreira and M.~Lagos,
  Phys.\ Rev.\ D {\bf 97} (2018) no.8,  084005.

\bibitem{Mukherjee}
S.~Mukherjee and S.~Chakraborty,
  Phys.\ Rev.\ D {\bf 97} (2018) no.12,  124007.

\bibitem{Chakra}
S.~Chakrabarti,
  Eur.\ Phys.\ J.\ C {\bf 78} (2018) no.4,  296.

\bibitem{Berti} E.~Berti, K.~Yagi and N.~Yunes,
  Gen.\ Rel.\ Grav.\  {\bf 50} (2018) no.4,  46.

\bibitem{Brihaye}
  Y.~Brihaye and B.~Hartmann,
  Class.\ Quant.\ Grav.\  {\bf 35} (2018) no.17,  175008.

\bibitem{Prabhu} K.~Prabhu and L.~C.~Stein,
  Phys.\ Rev.\ D {\bf 98} (2018) no.2,  021503.

\bibitem{Myung} Y.~S.~Myung and D.~C.~Zou,
  Phys.\ Rev.\ D {\bf 98} (2018) no.2,  024030;
  arXiv:1812.03604 [gr-qc].

\bibitem{Don-Kunz}
  J.~L.~Blazquez-Salcedo, D.~D.~Doneva, J.~Kunz and S.~S.~Yazadjiev,
  Phys.\ Rev.\ D {\bf 98} (2018) no.8,  084011;
J.~L.~Blazquez-Salcedo, Z.~Altaha Motahar, D.~D.~Doneva, F.~S.~Khoo, J.~Kunz, S.~Mojica, K.~V.~Staykov and S.~S.~Yazadjiev,
  arXiv:1810.09432 [gr-qc].

\bibitem{Benkel2018}
  R.~Benkel, N.~Franchini, M.~Saravani and T.~P.~Sotiriou,
  Phys.\ Rev.\ D {\bf 98} (2018) no.6,  064006.

\bibitem{Iorio} L.~Iorio and M.~L.~Ruggiero,
  JCAP {\bf 1810} (2018) no.10,  021.

\bibitem{Ovalle}
  J.~Ovalle, R.~Casadio, R.~da Rocha, A.~Sotomayor and Z.~Stuchlik,
  EPL {\bf 124} (2018) no.2,  20004;
  J.~Ovalle,
  Phys.\ Lett.\ B {\bf 788} (2019) 213.

\bibitem{Barack}
  L.~Barack {\it et al.},
   Class.\ Quant.\ Grav.\  {\bf 36} (2019) no.14,  143001.

\bibitem{Gao} Y.~X.~Gao, Y.~Huang and D.~J.~Liu,
  Phys.\ Rev.\ D {\bf 99} (2019) no.4,  044020.

\bibitem{Lee} B.~H.~Lee, W.~Lee and D.~Ro,
  Phys.\ Rev.\ D {\bf 99} (2019) no.2,  024002.

\bibitem{Witek2018}
  H.~Witek, L.~Gualtieri, P.~Pani and T.~P.~Sotiriou,
  Phys.\ Rev.\ D {\bf 99} (2019) no.6,  064035.

\bibitem{Moto} H.~Motohashi and S.~Mukohyama,
  Phys.\ Rev.\ D {\bf 99} (2019) no.4,  044030.

\bibitem{Kazanas}
  J.~Sultana and D.~Kazanas,
  Gen.\ Rel.\ Grav.\  {\bf 50} (2018) no.11,  137.

\bibitem{Nojiri-Odintsov-Oikonomou}
S.~Nojiri, S.~D.~Odintsov and V.~K.~Oikonomou,
Phys.\ Rev. \ D {\bf 99} (2019) 044050.

\bibitem{Cunha} 
  P.~V.~P.~Cunha, C.~A.~R.~Herdeiro and E.~Radu,
  Phys.\ Rev.\ Lett.\  {\bf 123}, no. 1, 011101 (2019).

\bibitem{Minamitsuji-Ikeda_Dec2018}
M.~Minamitsuji and T.~Ikeda,
Phys.\ Rev. \ D {\bf 99} (2019) 044017;
Phys.\ Rev. \ D {\bf 99} (2019) 104069.

\bibitem{Stetsko_Dec2018}
M.~M.~Stetsko,
Phys.\ Rev. \ D {\bf 99} (2019) 044028.

\bibitem{Myung_Dec2018}
Y.~S.~Myung and D.-C.~Zou,
Phys.\ Lett. \ B {\bf 790} (2019) 400.

\bibitem{Brihaye-Ducobu}
Y.~Brihaye and L.~Ducobu,
Phys.\ Lett. \ B {\bf 795} (2019) 135.

\bibitem{Herdeiro-Radu}
C.~A.~R.~Herdeiro and E.~Radu,
Phys.\ Rev. \ D {\bf 99} (2019) 084039.

\bibitem{Kobayashi}
T.~Kobayashi,
Rept.\ Prog. \ Phys. {\bf 82} (2019) 086901.

\bibitem{Macedo_Dec2018}
H.~O.~Silva, C.~F.~B.~Macedo, T.~P.~Sotiriou, L.~Gualtieri, J.~Sakstein and E.~Berti,
Phys.\ Rev. \ D {\bf 99} (2019) 104041.

\bibitem{Dombriz}
A.~de la Cruz-Dombriz and F.~J.M.~Torralba,
JCAP {\bf 1903} (2019) 002.

\bibitem{Wang-Shen-Xie}
C.-Y.~Wang, Y.-F.~Shen and Y.~Xie,
JCAP {\bf 1904} (2019) 022.

\bibitem{Cano-Ruiperez}
P.-A.~Cano and A.~Ruiperez,
JHEP {\bf 1905} (2019) 189.

\bibitem{Ramazanoglu_Jan2019}
F.~M.~Ramazanoglu,
Phys.\ Rev. \ D {\bf 99} (2019) 084015.

\bibitem{Fernandes}
P.~G.~S.~Fernandes, C.~A.~R.~Herdeiro, A.~M.~Pombo, E.~Radu and N.~Sanchis-Gual,
Class.\ Quant. \ Grav. {\bf 36} (2019) 134002.

\bibitem{Brihaye_Feb2019}
Y.~Brihaye and B.~Hartmann,
Phys. \ Lett. \ B {\bf 792} (2019) 244.

\bibitem{Saravani-Sotiriou}
M.~Saravani and T.~P.~Sotiriou,
Phys.\ Rev. \ D {\bf 99} (2019) 124004.

\bibitem{Macedo_Mar2019}
C.~F.~B.~Macedo, J.~Sakstein, E.~Berti, L.~Gualtieri, H.~O.~Silva and T.~P.~Sotiriou,
Phys.\ Rev. \ D {\bf 99} (2019) 104041.

\bibitem{Doneva-Staykov-Yazadjiev}
D.~D.~Doneva, K.~V.~Staykov and S.~S.~Yazadjiev,
Phys.\ Rev. \ D {\bf 99} (2019) 104045.

\bibitem{Saffer}
A.~Saffer, H.~O.~Silva and N.~Yunes,
Phys.\ Rev. \ D {\bf 100} (2019) 044030.

\bibitem{Anson}
T.~Anson, E.~Babichev, C.~Charmousis and S.~Ramazanov,
JCAP {\bf 1906} (2019) 023.

\bibitem{Myung_Mar2019}
Y.~S.~Myung and D.-C.~Zou,
Int.\ J. \ Mod. \ Phys. \ D {\bf 28} (2019) 1950114.

\bibitem{Brihaye-Hartmann_Mar2019}
Y.~Brihaye and B.~Hartmann,
JHEP {\bf 1909} (2019) 049.

\bibitem{Tattersall}
O.~J.~Tattersall and P.~G.~Ferreira,
Phys.\ Rev. \ D {\bf 99} (2019) 104082.

\bibitem{Andreou}
N.~Andreou, N.~Franchini, G.~Ventagli and T.~P.~Sotiriou,
Phys.\ Rev. \ D {\bf 99} (2019) 124022.

\bibitem{Liang-Sakstein-Trodden}
Q.~Liang, J.~Sakstein and M.~Trodden,
Phys.\ Rev. \ D {\bf 100} (2019) 063518.

\bibitem{Hui-Kabat}
L.~Hui, D.~Kabat, X.~Li, L.~Santoni and S.~S.~C.~Wong,
JCAP {\bf 1906} (2019) 038.

\bibitem{Tuan}
D.~Q.~Tuan and S.~H.~Q.~Nguyen,
Commun.\ Phys. {\bf 29} (2019) 173.

\bibitem{Tuan Do}
Tuan Do \textit{et al},
Science {\bf 365} (2019) 6454.

\bibitem{Fernandes-Herdeiro-Pombo-Radu-Gual_Jul2019}
P.~G.~S.~Fernandes, C.~A.~R.~Herdeiro, A.~M.~Pombo, E.~Radu and N.~Sanchis-Gual,
Phys.\ Rev. \ D {\bf 100} (2019) 084045.

\bibitem{Konoplya-Zhidenko}
R.~A.~Konoplya and A.~Zhidenko,
Phys.\ Rev. \ D {\bf 100} (2019) 044015.


\bibitem{Franchini-Sotiriou}
N.~Franchini and T.~P.~Sotiriou,
arXiv:1903.05427 [gr-qc].

\bibitem{Hees}
A.~Hees, O.~Minazzoli, E.~Savalle, Y.~V.~Stadnik, P.~Wolf and B.~Roberts,
arXiv:1905.08524 [gr-qc].

\bibitem{Anson-Babichev-Ramazanov}
T.~Anson, E.~Babichev and S.~Ramazanov,
arXiv:1905.10393 [gr-qc].

\bibitem{Khalil}
M.~Khalil, N.~Sennett, J.~Steinhoff and A.~Buonanno,
arXiv:1906.08161 [gr-qc].

\bibitem{Rham}
C.~de Rham and J.~Zhang,
arXiv:1907.006992 [hep-th].

\bibitem{Perez-Cruz-Lepe-Rivera}
G.~Aguilar-Perez, M.~Cruz, S.~Lepe and I.~Moran-Rivera,
arXiv:1907.06168 [gr-qc].

\bibitem{Konoplya-Pappas-Zhidenko}
R.~A.~Konoplya, T.~Pappas and A.~Zhidenko,
arXiv:1907.10112 [gr-qc].

\bibitem{Gao-Liu_Aug2019}
Y.-X.~Gao and D.-J.~Liu,
arXiv:1908.01346 [gr-qc].

\bibitem{Ikeda-Nakamura-Minamitsuji}
T.~Ikeda, T.~Nakamura and M.~Minamitsuji,
arXiv:1908.09394 [gr-qc].

\bibitem{Julie-Berti}
F.-L.~Julie and E.~Berti,
arXiv:1909.05258 [gr-qc].

\bibitem{Ramazanoglu_Oct2019}
F.~M.~Ramazanoglu and K.~I.~Unluturk,
arXiv:1910.02801 [gr-qc].

\bibitem{Aelst}
K.~V.~Aelst, E.~Gourgoulhon and P.~Grandclement,
arXiv:1910.08451 [gr-qc].

\bibitem{Barrientos}
J.~Barrientos, F.~Cordonier-Tello, C.~Corral, F.~Izaurieta, P.~Medina, E.~Rodriguez and O.~Valdivia,
arXiv:1910.00148 [gr-qc].




\bibitem{Martinez-deSitter}
  C.~Martinez, R.~Troncoso and J.~Zanelli,
  Phys.\ Rev.\ D {\bf 67} (2003) 024008.

\bibitem{Harper}
  T.~J.~T.~Harper, P.~A.~Thomas, E.~Winstanley and P.~M.~Young,
  Phys.\ Rev.\ D {\bf 70} (2004) 064023.

\bibitem{Martinez} M.~Henneaux, C.~Martinez, R.~Troncoso and J.~Zanelli,
  Phys.\ Rev.\ D {\bf 70} (2004) 044034;
  C.~Martinez, R.~Troncoso and J.~Zanelli,
  Phys.\ Rev.\ D {\bf 70} (2004) 084035;
  C.~Erices and C.~Martinez,
  Phys.\ Rev.\ D {\bf 97} (2018) no.2,  024034.

\bibitem{Radu-Win}
  E.~Radu and E.~Winstanley,
  Phys.\ Rev.\ D {\bf 72} (2005) 024017.

\bibitem{Anabalon}
  A.~Anabalon and H.~Maeda,
  Phys.\ Rev.\ D {\bf 81} (2010) 041501.

\bibitem{Hosler}
  D.~Hosler and E.~Winstanley,
  Phys.\ Rev.\ D {\bf 80} (2009) 104010.

\bibitem{Kolyvaris} C.~Charmousis, T.~Kolyvaris and E.~Papantonopoulos,
  Class.\ Quant.\ Grav.\  {\bf 26} (2009) 175012;
  T.~Kolyvaris, G.~Koutsoumbas, E.~Papantonopoulos and G.~Siopsis,
  Gen.\ Rel.\ Grav.\  {\bf 43} (2011) 163.

\bibitem{Ohta}
  K.~i.~Maeda, N.~Ohta and Y.~Sasagawa,
  Phys.\ Rev.\ D {\bf 83} (2011) 044051;
Z.~K.~Guo, N.~Ohta and T.~Torii,
  Prog.\ Theor.\ Phys.\  {\bf 121} (2009) 253;
N.~Ohta and T.~Torii,
  Prog.\ Theor.\ Phys.\  {\bf 121} (2009) 959; 
 N.~Ohta and T.~Torii,
  Prog.\ Theor.\ Phys.\  {\bf 122} (2009) 1477.

\bibitem{Saenz}
  S.~G.~Saenz and C.~Martinez,
  Phys.\ Rev.\ D {\bf 85} (2012) 104047.

\bibitem{Caldarelli}
  M.~M.~Caldarelli, C.~Charmousis and M.~Hassaine,
  JHEP {\bf 1310} (2013) 015.

\bibitem{Gonzalez}
  P.~A.~Gonzalez, E.~Papantonopoulos, J.~Saavedra and Y.~Vasquez,
  JHEP {\bf 1312} (2013) 021.

\bibitem{Gaete}
M.~Bravo Gaete and M.~Hassaine,
  Phys.\ Rev.\ D {\bf 88} (2013) 104011;
  M.~Bravo Gaete and M.~Hassaine,
  JHEP {\bf 1311} (2013) 177.

\bibitem{Giribet}
  G.~Giribet, M.~Leoni, J.~Oliva and S.~Ray,
  Phys.\ Rev.\ D {\bf 89} (2014) no.8,  085040.

\bibitem{BenAchour}
  J.~Ben Achour and H.~Liu,
  arXiv:1811.05369 [gr-qc].

\bibitem{Hartmann} Y.~Brihaye, B.~Hartmann and J.~Urrestilla,
  JHEP {\bf 1806} (2018) 074;
  Y.~Brihaye and B.~Hartmann,
  arXiv:1810.05108 [gr-qc].

\bibitem{BAK} A.~Bakopoulos, G.~Antoniou and P.~Kanti,
  Phys.\ Rev.\ D {\bf 99} (2019) no.6,  064003.

\bibitem{Achour-Liu}
J.~B.~Achour and H.~Liu,
Phys.\ Rev. \ D {\bf 99} (2019) 064042.

\bibitem{KBP_Corfu}
P.~Kanti, A.~Bakopoulos and N.~Pappas,
PoS \ CORFU2018 (2019) 091.

\bibitem{Radu-scal-dS}
Y.~Brihaye, C.~Herdeiro and E.~Radu,
arXiv:1910.05286 [gr-qc].


\bibitem{KKK1}
  P.~Kanti, B.~Kleihaus and J.~Kunz,
  Phys.\ Rev.\ Lett.\  {\bf 107} (2011) 271101;
%
  Phys.\ Rev.\ D {\bf 85} (2012) 044007.

\bibitem{ABKKK} 
  G.~Antoniou, A.~Bakopoulos, P.~Kanti, B.~Kleihaus and J.~Kunz,
  arXiv:1904.13091 [hep-th].


\bibitem{KKK2}
  B.~Kleihaus, J.~Kunz and P.~Kanti,
  arXiv:1910.02121 [gr-qc].

\bibitem{Herdeiro-Oliveira}
C.~A.~R.~Herdeiro and J.~M.~S.~Oliveira,
Class.\ Quant. \ Grav. {\bf 36} (2019) 105015.

\bibitem{Afonso}
  V.~I.~Afonso, G.~J.~Olmo, E.~Orazi and D.~Rubiera-Garcia,
  arXiv:1906.04623 [hep-th].

\bibitem{Radu-part}
  C.~A.~R.~Herdeiro, J.~M.~S.~Oliveira and E.~Radu,
  arXiv:1910.11021 [gr-qc].

\bibitem{Canate1}
P.~Canate, J.~Sultana and D.~Kazanas,
Phys.\ Rev. \ D {\bf 100} (2019) 064007.

\bibitem{Canate2}
P.~Canate and N.~Breton,
Phys.\ Rev. \ D {\bf 100} (2019) 064067.



\bibitem{ART} I.~Antoniadis, J.~Rizos and K.~Tamvakis,
  Nucl.\ Phys.\ B {\bf 415} (1994) 497.

\bibitem{RT}
  J.~Rizos and K.~Tamvakis,
  Phys.\ Lett.\ B {\bf 326} (1994) 57.

\bibitem{KRT} P.~Kanti, J.~Rizos and K.~Tamvakis,
  Phys.\ Rev.\ D {\bf 59} (1999) 083512.
  
\bibitem{KGD}
  P.~Kanti, R.~Gannouji and N.~Dadhich,
  Phys.\ Rev.\ D {\bf 92} (2015) no.4,  041302;
  Phys.\ Rev.\ D {\bf 92} (2015) no.8,  083524.



\end{thebibliography}
\end{document}